\documentclass[final,3p,times, onecolumn]{elsarticle}

\usepackage{latexsym}
\usepackage{bm}

\usepackage{amsmath}
\usepackage{graphicx}

\usepackage{amssymb}

\begin{document}

\begin{frontmatter}



\title{Introduction to Einstein-Maxwell equations and the Rainich conditions}


\author{Wytler Cordeiro dos Santos}
\ead{wytler@fis.unb.br}
\address{Universidade de
Bras\'\i lia, CEP 70910-900, DF, Brasil}

\begin{abstract}
The first results of Einstein-Maxwell equations established by Raincih in 1925 are therefore called the 
Raincih conditions. Later the result was rediscovered by Misner and Wheeler in 1957 and made the basis
of their geometrodynamics. The present survey will consider didactically 
the curvature  of spacetime attributed to
an electromagnetic field with conceptual and calculational details.
\end{abstract}

\begin{keyword}
The Rainich conditions, Einstein-Maxwell equations, electrovac and pure electromagnetic radiation



\end{keyword}

\end{frontmatter}


\section{Introduction}

The fields, such as the electromagnetic field describe the matter content of spacetime 
$({\cal M},\bm g)$, where ${\cal M}$ is a connected four-dimensional manifold and $\bm g$
is a Lorentz metric. These fields obey equations which can be expressed as relations between tensors
on manifold ${\cal M}$ in which all derivatives with respect to position are covariant derivatives
with respect to the symmetric connection
defined by the metric  $\bm g$. The equations governing the matter fields are such that there exist
a symmetric tensor $T_{\mu\nu}$, called the energy-momentum tensor, such that the relation between the
metric field $g_{\mu\nu}$ and the material contents of spacetime is expressed by 
Einstein's field equation \cite{Hawking},
\begin{equation}
\label{field equation}
R_{\mu\nu}-\frac{1}{2}g_{\mu\nu}R +\Lambda g_{\mu\nu}=\kappa T_{\mu\nu},
\end{equation}
$\Lambda$ is cosmological constant, $R_{\mu\nu}$ the contract curvature tensor (Ricci tensor), $R$ its trace
and $\kappa=\frac{8\pi G}{c^4}$ is a constant linked with the gravitational Newton's constant.
The energy-momentum symmetric tensor $T_{\mu\nu}$ depends on the fields, their covariant derivatives and the metric.

The {\it Einstein-Maxwell equations} of gravitation and electromagnetism consist of equation
(\ref{field equation}) (without cosmological constant),
with equations of motion $\nabla_{\mu}T^{\mu\nu}=0$, where appears in $T^{\mu\nu}$ the 
Maxwell energy-momentum tensor, and of the generally covariant Maxwell equations 
\cite{Rainich,Misner, Witten,Stephani}.

Greek labels refer to four dimensional spacetime; Latin labels refer to three dimensional space. 
The time coordinate receives the label 0, and we use metric signature (-+++).

\section{Electromagnetic Field in Minkowski Flat Spacetime}
The fields on Standard Model in the description of fundamental interactions, are fields in Minkowski
flat spacetime $(\mathbb{R}^4, \bm\eta)$, the spacetime of Special Relativity. The electromagnetism
is described by a vector potential $A^{\mu}=(\phi,\bm A)$, where the electromagnet field is obtained from a 
4-dimensional curl,
\begin{equation}
 \label{electromagnetic_tensor_1}
 F_{\mu\nu}= \partial_{\mu}A_{\nu} - \partial_{\nu}A_{\mu}.
\end{equation}
$F_{\mu\nu}$ is the  field-strength tensor of electromagnetism (electromagnetic field tensor).
Because of antisymmetry of electromagnetic field tensor, $F_{\mu\nu} = -F_{\nu\mu}$, the number
of independent terms is the number of combinations $\begin{pmatrix} 4\cr 2\end{pmatrix}=\dfrac{4!}{2!2!}=6$.
The antisymmetric field-strength tensor of electromagnetism has six components in all, $(E_x,E_y,E_z)$
and $(B_x,B_y,B_z)$. 
 
The tensor $F_{\mu\nu}$ has components (recall that  $A_{\mu}=(-\phi,\bm A)$)
\begin{equation}
 F_{i0}=\partial_i A_{0} - \partial_t A_i=\left [ \nabla (-\phi)-\frac{\partial {\bm A}}{\partial t}\right]_i
 =E_i,
\end{equation}
for $i=1,2,3$ we have three components: $(E_1,E_2,E_3)= (E_x,E_y,E_z)=\bm E$, of electric field. 
The others three components are obtained from
\begin{equation}
  F_{ij}=\partial_{i}A_{j} - \partial_{j}A_{i} \rightarrow(\nabla\times{\bm A})_k=B_k.
\end{equation}
For $i,j,k=1,2,3$ we have  $(B_1,B_2,B_3)= (B_x,B_y,B_z)=\bm B$.
Consider, for example, the component $F_{12}=\partial_1 A_2 - \partial_2 A_1=\partial_x A_y - \partial_y A_x=B_z$.
The antisymmetric field-strength tensor of electromagnetism may be displayed in matrix form, with the rows and
columns corresponding to the numbers 0,1,2,3:
\begin{equation}
 (F_{\mu\nu})=
 \begin{pmatrix}
  0 & -E_x & -E_y & -E_z \cr
   E_x & 0 & B_z & -B_y \cr
   E_y & -B_z & 0 & B_x \cr
   E_z & B_y & -B_x & 0 \cr
 \end{pmatrix}.
\end{equation}
The electromagnetic field tensor with two contravariants indices is obtained below
\begin{equation}
 (F^{\mu\nu})=(\eta^{\mu\rho}F_{\rho\sigma}\eta^{\sigma\nu})=
 \begin{pmatrix}
  0 & E_x & E_y & E_z \cr
  -E_x & 0 & B_z & -B_y \cr
  -E_y & -B_z & 0 & B_x \cr
  -E_z & B_y & -B_x & 0 \cr
 \end{pmatrix}.
\end{equation}
The elements of  $(F^{\mu\nu})$ are obtained from $ (F_{\mu\nu})$ by putting
${\bm E}\rightarrow -{\bm E}$. 
 
\subsection{The dual tensor of electromagnetism} 

In Classical Field Theory it is convenient and useful
to define the dual tensor with the aid of a pseudotensor,
\begin{equation}
\label{dual_1}
 \tilde{F}_{\mu\nu}=\frac{1}{2} \epsilon_{\rho\sigma\mu\nu}F^{\rho\sigma} 
 = \frac{1}{2} \epsilon_{\mu\nu\rho\sigma}F^{\rho\sigma},
\end{equation}
where $\epsilon_{\rho\sigma\mu\nu}$ is the Levi-Civita symbol in four dimensions, with $\epsilon_{0123}=-1$
and totally antisymmetric with respect to all pairs of indices.
One can obtain the components of dual tensor with the following calculations:
\begin{eqnarray}
 \tilde{F}_{01} &=&\frac{1}{2}\epsilon_{01ij}F^{ij}= \frac{1}{2}\epsilon_{0123}F^{23}+
 \frac{1}{2}\epsilon_{0132}F^{32}\cr
 &=& \frac{1}{2}\epsilon_{0123}F^{23}+
 \frac{1}{2}(-\epsilon_{0123})(-F^{23})= \epsilon_{0123}F^{23},\nonumber
\end{eqnarray}
where $\epsilon_{0132}=-\epsilon_{0123}$ (odd permutation) and $F^{23}=-F^{32}$, that results:
\begin{equation}
  \tilde{F}_{01} = - B_x .\nonumber
\end{equation}
Similarly for $\tilde{F}_{02}$ we have:
\begin{eqnarray}
 \tilde{F}_{02} &=& \frac{1}{2}\epsilon_{0213}F^{13}+
 \frac{1}{2}\epsilon_{0231}F^{31}=\epsilon_{0213}F^{13}=(+1)(-B_y)= -B_y\nonumber
\end{eqnarray}
and for $\tilde{F}_{03}$ :
\begin{eqnarray}
 \tilde{F}_{03} &=& \frac{1}{2}{\epsilon}_{0312}F^{12}+
 \frac{1}{2}\epsilon_{0321}F^{21}=\epsilon_{0312}F^{12}=(-1)B_z= -B_z.\nonumber
\end{eqnarray}
The others three components are:
\begin{eqnarray}
  \tilde{F}_{12} &=&  \epsilon_{1203}F^{03}=-E_z\cr
  \tilde{F}_{13} &=&  \epsilon_{1302}F^{02}=+E_y\cr
   \tilde{F}_{23} &=&  \epsilon_{2301}F^{01}=-E_x.\nonumber
\end{eqnarray}
Explicitly, in matrix form,
\begin{equation}
\label{dual_F_2}
 (\tilde{F}_{\mu\nu})=
 \begin{pmatrix}
  0 & -B_x & -B_y & -B_z \cr
  B_x & 0 & -E_z & E_y \cr
  B_y & E_z & 0 & -E_x \cr
  B_z & -E_y & E_x & 0 \cr
 \end{pmatrix}
\end{equation}
The change $F_{\mu\nu} \rightarrow \tilde{F}_{\mu\nu}$ is obtained by putting 
${\bm E} \rightarrow {\bm B}$ e ${\bm B} \rightarrow -{\bm E}$.

The Maxwell equations are in Lorentz covariant form. Two inhomogeneous equations are contained
in the covariant equation
\begin{equation}
\label{Maxwell_equations_1}
 \partial_{\mu}F^{\mu\nu} = -j^{\nu},
\end{equation}
where $j^{\nu} = (\rho,\bm j)$ is the 4-current, that yields:
\begin{equation}
 \nabla\cdot{\bm E} = \rho \hspace*{1cm}\mbox{and}\hspace*{1cm} 
 \nabla\times{\bm B} -\frac{\partial{\bm E}}{\partial t} = {\bm j}.\nonumber
\end{equation}

The two homogeneous Maxwell equations are obtained from dual field-strength tensor (\ref{dual_F_2}) as
\begin{equation}
\label{Maxwell_equations_1.1}
 \partial_{\mu}\tilde{F}^{\mu\nu} = 0,
\end{equation}
yielding
\begin{equation}
 \nabla\cdot{\bm B} =0 \hspace*{1cm}\mbox{and}\hspace*{1cm} 
 \nabla\times{\bm E} + \frac{\partial{\bm E}}{\partial t} = 0.\nonumber
\end{equation}
Consider, for example, $\nu=0$,
\begin{equation}
 \partial_{i}\tilde{F}^{i0} =  \partial_{1}\tilde{F}^{10}+\partial_{2}\tilde{F}^{20}+\partial_{3}\tilde{F}^{30}
 = \nabla\cdot{\bm B} =0.
\end{equation}

The equation (\ref{Maxwell_equations_1.1}), $\partial_{\mu}\tilde{F}^{\mu\nu} = 0$, can be written in another way
by contraction $\epsilon_{\rho\sigma\tau\nu} \partial_{\mu}\tilde{F}^{\mu\nu} = 0$, that in this case it becomes:
\begin{eqnarray}
 \epsilon_{\rho\sigma\tau\nu} \partial_{\mu}(\epsilon^{\mu\nu\kappa\lambda}F_{\kappa\lambda}) =
  \epsilon_{\rho\sigma\tau\nu}\epsilon^{\mu\nu\kappa\lambda}\partial_{\mu}F_{\kappa\lambda}= 0\cr
  -({\delta^{\kappa}}_{\rho}{\delta^{\lambda}}_{\sigma} {\delta^{\mu}}_{\tau}+
 {\delta^{\kappa}}_{\sigma}{\delta^{\lambda}}_{\tau} {\delta^{\mu}}_{\rho} +
 {\delta^{\kappa}}_{\tau}{\delta^{\lambda}}_{\rho} {\delta^{\mu}}_{\sigma} -
 {\delta^{\kappa}}_{\tau}{\delta^{\lambda}}_{\sigma} {\delta^{\mu}}_{\rho} -
  {\delta^{\kappa}}_{\sigma}{\delta^{\lambda}}_{\rho} {\delta^{\mu}}_{\tau} -
  {\delta^{\kappa}}_{\rho}{\delta^{\lambda}}_{\tau} {\delta^{\mu}}_{\sigma})\partial_{\mu}F_{\kappa\lambda}=0,
  \nonumber
\end{eqnarray}
that reduces:
\begin{equation}
\partial_{\tau}F_{\rho\sigma} + \partial_{\rho}F_{\sigma\tau} + \partial_{\sigma}F_{\tau\rho} = 0.
\end{equation}

\subsection{Bivectors}

Bivectors are 'antisymmetric tensors` of second order. A simple bivector,
\begin{equation}
 X_{\kappa\lambda} = u_{\kappa}v_{\lambda} - u_{\lambda}v_{\kappa},
\end{equation}
represents a 2-surface element spanned by two tangent vectors $u_{\kappa}$ and $v_{\lambda}$.
Similarly to vectors, this surface element is spacelike, timelike or null according as 
$X_{\kappa\lambda} X^{\kappa\lambda}$ is positive, negative or zero, respectively. As a result, the components
of electromagnetic field tensor, $F_{\mu\nu}$ are coordinates of a bivector in Minkowski flat spacetime. Bivectors
satisfy some useful properties which follow below:

\begin{itemize}
 \item Let us return to duality operation, and calculate a 
repeated application of the duality  operation, $\widetilde{\widetilde{X}}_{\mu\nu}$, where
$\widetilde{X}^{\kappa\lambda}=\frac{1}{2}\epsilon^{\kappa\lambda\rho\sigma}X_{\rho\sigma}$.
gives
\begin{eqnarray}
 \widetilde{\widetilde{X}}_{\mu\nu} = \frac{1}{2}\epsilon_{\mu\nu\kappa\lambda}\widetilde{X}^{\kappa\lambda} =
\frac{1}{2}\epsilon_{\mu\nu\kappa\lambda}\,\,\cdot\,\,
 \frac{1}{2}\epsilon^{\kappa\lambda\rho\sigma}X_{\rho\sigma}\nonumber
\end{eqnarray}
with the contraction of the Levi-Civita tensors, we have
\begin{eqnarray}
 \widetilde{\widetilde{X}}_{\mu\nu} = 
 \frac{1}{4} (-2)({\delta_{\mu}}^{\rho}{\delta_{\nu}}^{\sigma}-{\delta_{\mu}}^{\sigma}{\delta_{\nu}}^{\rho} )
 X_{\rho\sigma}
  = -\frac{1}{2}(X_{\mu\nu} - X_{\nu\mu}),\nonumber
\end{eqnarray}
or
\begin{equation}
\label{dual_dual_bivector}
  \widetilde{\widetilde{X}}_{\mu\nu} = -  X_{\nu\mu}.
\end{equation}
\item Two bivectors $X_{\mu\nu}$ and $Y_{\mu\nu}$ satisfy the identity
\begin{equation}
\widetilde{X}_{\mu\nu} Y^{\mu\nu} = \left(\frac{1}{2}\epsilon_{\mu\nu\rho\sigma}X^{\rho\sigma}\right) Y^{\mu\nu}
=X^{\rho\sigma}\left( \frac{1}{2}\epsilon_{\rho\sigma\mu\nu}Y^{\mu\nu}\right)=
X^{\rho\sigma}\widetilde{Y}_{\rho\sigma}\nonumber
\end{equation}
or 
\begin{equation}
\label{identidade_1}
\widetilde{X}_{\mu\nu} Y^{\mu\nu} = X_{\mu\nu} \widetilde{Y}^{\mu\nu}.
\end{equation}
\item Let $X_{\mu\nu}$ and $Y_{\mu\nu}$ bivectors such that
\begin{eqnarray}
 \widetilde{X}_{\mu\rho}\widetilde{Y}^{\nu\rho} &=& \frac{1}{2}\epsilon_{\mu\rho\sigma\tau}X^{\sigma\tau}
 \cdot \frac{1}{2}\epsilon^{\nu\rho\phi\chi}Y_{\phi\chi}\cr
 &=& \frac{1}{4}  X^{\sigma\tau}Y_{\phi\chi}     \epsilon_{\mu\sigma\tau\rho}\epsilon^{\nu\phi\chi\rho}\cr
 &=& \frac{1}{4}  X^{\sigma\tau}Y_{\phi\chi}  
 (-1)({\delta_{\mu}}^{\nu}{\delta_{\sigma}}^{\phi}{\delta_{\tau}}^{\chi} - 
 {\delta_{\mu}}^{\nu}{\delta_{\sigma}}^{\chi}{\delta_{\tau}}^{\phi} +
 {\delta_{\mu}}^{\chi}{\delta_{\sigma}}^{\nu}{\delta_{\tau}}^{\phi} -
  {\delta_{\mu}}^{\chi}{\delta_{\sigma}}^{\phi}{\delta_{\tau}}^{\nu}\cr
  & & + {\delta_{\mu}}^{\phi}{\delta_{\sigma}}^{\chi}{\delta_{\tau}}^{\nu} -
   {\delta_{\mu}}^{\phi}{\delta_{\sigma}}^{\nu}{\delta_{\tau}}^{\chi})\cr
   &=& -\frac{1}{4}({\delta_{\mu}}^{\nu} X^{\sigma\tau}Y_{\sigma\tau}  - 
   {\delta_{\mu}}^{\nu} X^{\sigma\tau}Y_{\tau\sigma}+ X^{\nu\tau}Y_{\tau\mu}- X^{\sigma\nu}Y_{\sigma\mu}
   +X^{\sigma\nu}Y_{\mu\sigma}
    -X^{\nu\tau}Y_{\mu\tau})\cr
   &=&  -\frac{1}{4}({2 \delta_{\mu}}^{\nu} X^{\sigma\tau}Y_{\sigma\tau}  - 4  X^{\nu\tau} Y_{\mu\tau} ),\nonumber
\end{eqnarray}
that we end up with
\begin{equation}
\label{identidade_2}
 X_{\nu\rho} {Y_{\mu}}^{\rho} - \widetilde{X}_{\mu\rho}{\widetilde{Y}_{\nu}}\,^{\rho} =
 \frac{1}{2}\eta_{\mu\nu} X^{\sigma\tau}Y_{\sigma\tau}.
\end{equation}
Observe that
\begin{equation}
\label{identidade_3}
 \widetilde{X}_{\mu\nu}\widetilde{Y}^{\mu\nu} =  -{X}_{\mu\nu}{Y}^{\mu\nu}.
\end{equation}

\item If we put ${Y_{\mu}}^{\rho} =  {\widetilde{X}_{\mu}}\,^{\rho}$ in the identity (\ref{identidade_2}),
we can obtain other necessary identity,
\begin{eqnarray}
 X_{\nu\rho} {\widetilde{X}_{\mu}}\,^{\rho} - \widetilde{X}_{\mu\rho}{\widetilde{\widetilde{X}}_{\nu}}\,^{\rho} &=&
 \frac{1}{2}\eta_{\mu\nu} X^{\sigma\tau}{\widetilde{X}}_{\sigma\tau}\cr
 X_{\nu\rho} {\widetilde{X}_{\mu}}\,^{\rho} - \widetilde{X}_{\mu\rho}(-{X_{\nu}}^{\rho}) &=&
 \frac{1}{2}\eta_{\mu\nu} X^{\sigma\tau}{\widetilde{X}}_{\sigma\tau}\cr
 2 X_{\nu\rho} {\widetilde{X}_{\mu}}\,^{\rho} &=&
 \frac{1}{2}\eta_{\mu\nu} X^{\sigma\tau}{\widetilde{X}}_{\sigma\tau}\nonumber
\end{eqnarray}
or 
\begin{equation}
 \label{identidade_5}
 X_{\nu\rho} {\widetilde{X}_{\mu}}\,^{\rho} =
 \frac{1}{4}\eta_{\mu\nu} X^{\sigma\tau}{\widetilde{X}}_{\sigma\tau}.
\end{equation}

\end{itemize}

A complex bivector can be defined by:
\begin{equation}
 \label{bivetor_complexo_1}
 {\cal X}_{\mu\nu} = X_{\mu\nu}+ i \widetilde{X}_{\mu\nu}
\end{equation}
such that it is self-dual, i.e. it fulfills the condition:
\begin{eqnarray}
 \widetilde{\cal X}_{\mu\nu} &=& \frac{1}{2} \epsilon_{\mu\nu\rho\sigma}{\cal X}^{\rho\sigma}\cr
 &=&\frac{1}{2} \epsilon_{\mu\nu\rho\sigma}{X}^{\rho\sigma}+\frac{i}{2} 
 \epsilon_{\mu\nu\rho\sigma}\widetilde{X}^{\rho\sigma}\cr
 &=& \widetilde{X}_{\mu\nu} +i \widetilde{\widetilde{X}}_{\mu\nu}\cr
 &=& \widetilde{X}_{\mu\nu} +i (-{X}_{\mu\nu}),\nonumber
\end{eqnarray}
putting the imaginary number in evidence
\begin{equation}
 \label{bivetor_complexo_2}
\widetilde{\cal X}_{\mu\nu}  = -i({X}_{\mu\nu} + i  \widetilde{X}_{\mu\nu})= -i {\cal X}_{\mu\nu}.
\end{equation}


\subsection{Electromagnetic self-dual  bivector }

With aid of a timelike unit vector $u_{\mu}$, a self-dual bivector of electromagnetic field,
\begin{equation}
 \label{bivetor_complexo_3}
 {\cal F}_{\mu\nu} = F_{\mu\nu}+ i \widetilde{F}_{\mu\nu},
\end{equation}
is 
determined by the projection
\begin{equation}
\label{auto-dual_1}
 \mbox{F}_{\mu}={\cal F}_{\mu\nu}u^{\nu},
\end{equation}
where $u_{\nu}u^{\nu}=-1$. As result we have 
the orthogonality condition $  \mbox{F}_{\mu}u^{\mu}={\cal F}_{\mu\nu}u^{\nu}u^{\mu}=0$.
Let us write the electromagnetic field in terms of the projection $\mbox{F}_{\mu}$ and of  timelike unit 
vector $u_{\mu}$. The self-dual bivector of electromagnetic field (\ref{bivetor_complexo_3}) has real part
and  dual imaginary part defined by
\begin{equation}
\label{bivetor_complexo_4}
 {\cal F}_{\mu\nu} = u_{\mu}\mbox{F}_{\nu} - u_{\nu}\mbox{F}_{\mu}+
 \frac{i}{2} \epsilon_{\mu\nu\rho\sigma}( u^{\,\rho}\mbox{F}^{\sigma} - u^{\sigma}\mbox{F}^{\,\rho} ).
\end{equation}
The projection (\ref{auto-dual_1}) is obtained evaluating with orthogonality condition $  \mbox{F}_{\mu}u^{\mu}=0$
\begin{equation}
 {\cal F}_{\mu\nu}u^{\nu} =  u_{\mu}u^{\nu}\mbox{F}_{\nu} - u^{\nu}u_{\nu}\mbox{F}_{\mu}+
  i\epsilon_{\mu\nu\rho\sigma} u^{\nu} u^{\,\rho}\mbox{F}^{\sigma} = 0 -(-1)\mbox{F}_{\mu}+0=\mbox{F}_{\mu}. 
  \nonumber
\end{equation}

In an inertial frame in which it is at rest, the four-velocity is $u^{\mu} =(1,0,0,0)$. If we project the 
electromagnetic field-strength $F_{\mu\nu}$ in the direction of the vector $u^{\mu}$, we have
\begin{equation}
\label{campo_eletrico_1}
 F_{\mu\nu}u^{\nu} = \begin{pmatrix}
  0 & -E_x & -E_y & -E_z \cr
   E_x & 0 & B_z & -B_y \cr
   E_y & -B_z & 0 & B_x \cr
   E_z & B_y & -B_x & 0 \cr
 \end{pmatrix} \begin{pmatrix} 1 \cr 0 \cr 0\cr 0 
 \end{pmatrix} = \begin{pmatrix} 0 \cr E_x \cr E_y \cr E_z 
 \end{pmatrix},
\end{equation}
where we can denote this by: $F_{\mu\nu}u^{\nu} = E_{\mu}$. Similarly we can project the dual tensor of electromagnetic
field, ${\tilde{F}}_{\mu\nu}$ in the direction of the vector $u^{\mu}$,
\begin{equation}
\label{campo_magnetico_1}
 {\tilde{F}}_{\mu\nu}u^{\nu} = \begin{pmatrix}
  0 & -B_x & -B_y & -B_z \cr
  B_x & 0 & -E_z & E_y \cr
  B_y & E_z & 0 & -E_x \cr
  B_z & -E_y & E_x & 0 \cr
 \end{pmatrix} \begin{pmatrix} 1 \cr 0 \cr 0\cr 0 
 \end{pmatrix} = \begin{pmatrix} 0 \cr B_x \cr B_y \cr B_z 
 \end{pmatrix},
\end{equation}
where we can denote this by: ${\tilde{F}}_{\mu\nu}u^{\nu} = B_{\mu}$. Hence we can obtain that:
\begin{equation}
\label{vetor_eletromagnetico_complexo_1}
 \mbox{F}_{\mu} = E_{\mu}+iB_{\mu}.
\end{equation}
We can rewrite equation (\ref{bivetor_complexo_4}) as:
\begin{equation}
 {\cal F}_{\mu\nu} =  u_{\mu}(E_{\nu}+iB_{\nu}) - u_{\nu}(E_{\mu}+iB_{\mu})+
  i\epsilon_{\mu\nu\rho\sigma} u^{\,\rho}(E^{\sigma}+iB^{\sigma}),\nonumber
\end{equation}
where this can be separated in real and imaginary parts:
\begin{equation}
 {\cal F}_{\mu\nu} =  (u_{\mu}E_{\nu} - u_{\nu}E_{\mu} - \epsilon_{\mu\nu\rho\sigma}u^{\,\rho}B^{\sigma})
 +i  (u_{\mu}B_{\nu} - u_{\nu}B_{\mu} + \epsilon_{\mu\nu\rho\sigma}u^{\,\rho}E^{\sigma})\nonumber
\end{equation}
We have from (\ref{bivetor_complexo_3}), that  ${\cal F}_{\mu\nu}$ 
with its real and imaginary parts, 
$ {\cal F}_{\mu\nu}=F_{\mu\nu} + i \widetilde{F}_{\mu\nu} $,   yields the following terms:
\begin{equation}
\label{campo_Eletrico}
 F_{\mu\nu}= u_{\mu}E_{\nu} - u_{\nu}E_{\mu} - \epsilon_{\mu\nu\rho\sigma}u^{\,\rho}B^{\sigma}
\end{equation}
and
\begin{equation}
\label{campo_Magnetico}
 \widetilde{F}_{\mu\nu} = u_{\mu}B_{\nu} - u_{\nu}B_{\mu} + \epsilon_{\mu\nu\rho\sigma}u^{\,\rho}E^{\sigma}.
\end{equation}
The projections
\begin{equation}
\label{campo_eletrico_2}
 E_{\mu} = F_{\mu\nu}u^{\nu} = (0, E_x,E_y,E_z)
\end{equation}
and
\begin{equation}
\label{campo_magnetico_2}
  B_{\mu} = \widetilde{F}_{\mu\nu} u^{\nu} = (0, B_x,B_y,B_z),
\end{equation}
respectively denote the electric and magnetic parts of the electromagnetic field tensor $F_{\mu\nu}$.

As has been previously defined ${\cal F}_{\mu\nu}$ from (\ref{bivetor_complexo_3}), one can put 
inhomogeneous and homogeneous Maxwell equations as follows:
\begin{equation}
\label{Maxwell_equations_5}
 \partial_{\mu}{\cal F}^{\mu\nu} = -j^{\nu}.
\end{equation}
In fact, in the inertial frame in which it is at rest, the vectors $E_{\mu}$ and $B_{\mu}$ from 
(\ref{campo_eletrico_1}) and (\ref{campo_magnetico_1}) respectively, are spatial 3-vectors, where we can put 
\begin{equation}
\label{vetor_eletromagnetico_complexo_2}
 \mbox{\bf F} = {\bm E} + i{\bm B},
\end{equation}
Thus, we can use above complex 3-vector to put the four Maxwell equations into other form as follows:
\begin{equation}
 \nabla \cdot  \mbox{\bf F} = \rho \hspace*{1cm} \Rightarrow \hspace*{1cm}
 \begin{cases}
  \nabla \cdot {\bm E} =\rho \cr
  \nabla \cdot {\bm B} = 0 
 \end{cases}
\end{equation}
and
\begin{equation}
 -i\nabla \times  \mbox{\bf F} -\frac{\partial  \mbox{\bf F}}{\partial t} = {\bm j}\hspace*{1cm}  \Rightarrow \hspace*{1cm}
 \begin{cases}
  \nabla \times {\bm B} -\frac{\partial {\bm E}}{\partial t} = {\bm j}  \cr
  \nabla \times {\bm E} +\frac{\partial {\bm B}}{\partial t}  = 0 
 \end{cases}.
\end{equation}

\subsection{Lorentz transformations and invariants}
The laws of Physics  in the Minkowski spacetime are invariant under Lorentz transformations, 
that are linear and performing transformations between the coordinates of an
inertial frame ${\cal O}$ and other inertial frame ${\cal O}'$. Thus the transformation between two inertial
coordinate system for 4-vectors is given by:
\begin{equation}
\label{transformacoes_Lorentz_1}
 U'^{\mu} = {\Lambda^{\mu}}_{\nu} U^{\nu}.
\end{equation}
The magnitude of a vector in the Minkowski spacetime is a frame-independent scalar number under Lorentz transformation:
\begin{equation}
\label{invariacia_1}
 U'^2 =  U'^{\mu}U'_{\mu} = ({\Lambda^{\mu}}_{\rho} U^{\rho})({\Lambda_{\mu}}^{\sigma}U_{\sigma})
 = U^{\rho}{\delta_{\rho}}^{\sigma} U_{\sigma} = U^{\rho}U_{\rho} = U^2.
\end{equation}

Since the fields ${\bm E}$ and ${\bm B}$ are the elements of the electromagnetic filed tensor $F_{\mu\nu}$,
their values in one inertial frame ${\cal O}$ can be expressed 
in another inertial frame ${\cal O}'$ according to
\begin{equation}
 F'^{\mu\nu} ={\Lambda^{\mu}}_{\rho}{\Lambda^{\nu}}_{\sigma}F^{\rho\sigma}.
\end{equation}
In the same way the contraction $F^{\mu\nu}F_{\mu\nu}$ is a frame-independent scalar number 
under Lorentz transformation. We can see that the self-dual bivector 
of electromagnetic field is invariant,
\begin{equation}
  {\cal F}'_{\mu\nu} {\cal F}'^{\mu\nu} = ({\Lambda_{\mu}}^{\rho}{\Lambda_{\nu}}^{\sigma} {\cal F}_{\rho\sigma})
   ({\Lambda_{\kappa}}^{\mu}{\Lambda_{\lambda}}^{\nu} {\cal F}_{\kappa\lambda})=
   {\delta_{\kappa}}^{\rho}{\delta_{\lambda}}^{\sigma}{\cal F}_{\rho\sigma}  {\cal F}_{\kappa\lambda}= 
   {\cal F}_{\rho\sigma}  {\cal F}^{\rho\sigma}  \nonumber
\end{equation}
and from  (\ref{bivetor_complexo_3}) it follows that
\begin{equation}
  {\cal F}_{\mu\nu} {\cal F}^{\mu\nu}=( F_{\mu\nu}+ i \widetilde{F}_{\mu\nu})(F^{\mu\nu}+ i\widetilde{F}^{\mu\nu}) = 
  F_{\mu\nu}F^{\mu\nu} +i F_{\mu\nu}\widetilde{F}^{\mu\nu} + i \widetilde{F}_{\mu\nu}F^{\mu\nu}
  - \widetilde{F}_{\mu\nu}\widetilde{F}^{\mu\nu}, \nonumber
\end{equation}
where we can use (\ref{identidade_1}) and (\ref{identidade_3}) to obtain
\begin{equation}
  {\cal F}_{\mu\nu} {\cal F}^{\mu\nu} =
  2  F_{\mu\nu}F^{\mu\nu} + 2i F_{\mu\nu}\widetilde{F}^{\mu\nu}.
\end{equation}
The above expression
yields two Lorentz invariants $I_1$ and $I_2$ as follow:
\begin{itemize}
\item $I_1 = \dfrac{1}{2} F_{\mu\nu}F^{\mu\nu} $

A simplified expression for this Lorentz invariant can be obtained from (\ref{campo_Eletrico}),
\begin{eqnarray}
F_{\mu\nu}F^{\mu\nu} &=& (u_{\mu}E_{\nu} - u_{\nu}E_{\mu} - \epsilon_{\mu\nu\rho\sigma}u^{\rho}B^{\sigma})
(u^{\mu}E^{\nu} - u^{\nu}E^{\mu} - \epsilon^{\mu\nu\kappa\lambda}u_{\kappa}B_{\lambda})\cr
&=&u_{\mu}u^{\mu}E^{\nu}E_{\nu}-u_{\mu}E_{\nu} u^{\nu}E^{\mu} - 
\epsilon^{\mu\nu\kappa\lambda}u_{\mu}u_{\kappa}B_{\lambda}E_{\nu} - u_{\nu}E_{\mu}u^{\mu}E^{\nu}
+u_{\nu} u^{\nu}E^{\mu} E_{\mu}\cr
& & + \epsilon^{\mu\nu\kappa\lambda}u_{\nu}u_{\kappa}B_{\lambda}E_{\mu}
- \epsilon_{\mu\nu\rho\sigma}u^{\mu}u^{\rho}B^{\sigma}E^{\nu}
+ \epsilon_{\mu\nu\rho\sigma}u^{\nu} u^{\rho}B^{\sigma}E^{\mu}
+\epsilon_{\mu\nu\rho\sigma}\epsilon^{\mu\nu\kappa\lambda}u^{\rho}B^{\sigma}u_{\kappa}B_{\lambda},\nonumber
\end{eqnarray}
where we have that $u_{\mu}E^{\mu}=0$, $\epsilon_{\mu\nu\rho\sigma}u^{\mu}u^{\rho}=0$ and
 $u_{\mu}u^{\mu}=-1$ such as
\begin{eqnarray}
F_{\mu\nu}F^{\mu\nu} &=& - E^{\nu}E_{\nu} - E^{\mu}E_{\mu} -2({\delta_{\rho}}^{\kappa}{\delta_{\sigma}}^{\lambda}
-{\delta_{\rho}}^{\lambda}{\delta_{\sigma}}^{\kappa})u_{\kappa}B_{\lambda}u^{\rho}B^{\sigma}\cr
&=& -2 E^{\mu}E_{\mu} - 2(-B_{\sigma}B^{\sigma}),\nonumber
\end{eqnarray}
that in terms of (\ref{campo_eletrico_2}) and (\ref{campo_magnetico_2}) we have that the Lorentz
invariant is given by in terms of 3-vectors $\bm E$ and $\bm B$:
\begin{equation}
 F_{\mu\nu}F^{\mu\nu} = - 2 |{\bm E}|^2 + 2|{\bm B}|^2
\end{equation}
or
\begin{equation}
\label{invariante_1}
 I_1 = \frac{1}{2} F_{\mu\nu}F^{\mu\nu} = |{\bm B}|^2 - |{\bm E}|^2 .
\end{equation}

\item $I_ 2 = \dfrac{1}{2} F_{\mu\nu}\widetilde{F}^{\mu\nu} $

Similarly, from (\ref{campo_Eletrico}) and (\ref{campo_Magnetico}) and in terms of 3-vectors
$\bm E$ e $\bm B$, we have:
\begin{equation}
 F_{\mu\nu}\widetilde{F}^{\mu\nu} = -4{\bm E}\cdot {\bm B}
\end{equation}
or 
\begin{equation}
\label{invariante_2}
 I_2 = \frac{1}{2} {\widetilde F}_{\mu\nu}F^{\mu\nu} = -2{\bm E}\cdot {\bm B}.
\end{equation}
\end{itemize}

In Minkowski space field theory, the spin of a field can be classified according to the field's properties
under infinitesimal Lorentz transformations. A vector potential $A^{\mu}$ transforms from Lorentz frame
system ${\cal O}$ to other ${\cal O}'$ in accordance with (\ref{transformacoes_Lorentz_1}). Under infinitesimal
Lorentz transformation we have \cite{Blagojevic,Weinberg}:
\begin{equation}
 A'^{\mu} = \left\{{\delta^{\mu}}_{\nu} + \frac{i}{2}\omega^{\rho\sigma}{\left[\Sigma_{\rho\sigma}\right]^{\mu}}_{\nu}
\right\}A^{\nu},
\end{equation}
where the antisymmetric tensor, $\omega^{\,\rho\sigma}=-\omega^{\sigma\rho}$, are infinitesimal parameters, and 
the $\Sigma_{\rho\sigma}$ are  Hermitian generators, that satisfy a Lie algebra 
\begin{equation}
\label{algebra_Lie}
 i[\Sigma_{\kappa\lambda}, \Sigma_{\mu\nu}] = \eta_{\mu\lambda}\Sigma_{\kappa\nu} -
 \eta_{\mu\kappa}\Sigma_{\lambda\nu} + \eta_{\nu\lambda}\Sigma_{\mu\kappa} -\eta_{\nu\kappa}\Sigma_{\mu\lambda}.
\end{equation}
These generators can be identified with the group $SO(1,3)$ of Lorentz transformations. The generators 
of vector field are given by:
\begin{equation}
\label{gerador_lorentz}
 \Sigma_{\rho\sigma} \rightarrow {[\Sigma_{\rho\sigma}]^{\mu}}_{\nu} = 
  i\left( \eta_{\rho\nu}{\delta_{\sigma}}^{\mu} - \eta_{\sigma\nu}{\delta_{\rho}}^{\mu}\right).
\end{equation}
In Field Theory, the  vector potential $A^{\mu}$ that describes the electromagnetic field, can be classified with an 
irreducible representation of the Lorentz Group, $SO(1,3)$, where this vector field is massless and spin 1.

For  vector field, the infinitesimal Lorentz transformations results in:
\begin{equation}
 A'^{\mu} = A^{\nu}+ {\omega^{\mu}}_{\nu} A^{\nu}.
\end{equation}

For a general Lorentz transformation from inertial system ${\cal O}$ to a system ${\cal O}'$
moving with 3-vector velocity $\bm v$ relative to  ${\cal O}$, the transformation of the fields 
${\bm E}$ and ${\bm B}$ can be written \cite{Jackson}
\begin{eqnarray}
\label{transformacoes_Lorentz_E_B}
 {\bm E}' = \gamma\left( {\bm E}+{\bm v}\times {\bm B}\right) - \frac{\gamma^2}{\gamma +1} 
 {\bm v}\left({\bm v}\cdot {\bm E}\right)\cr
  {\bm B}' = \gamma\left( {\bm B}-{\bm v}\times {\bm E}\right) - \frac{\gamma^2}{\gamma +1} 
  {\bm v}\left({\bm v}\cdot {\bm B}\right),
\end{eqnarray}
where these transformations show that $\bm E$ and $\bm B$ have no independent existence.
If we choose an inertial system ${\cal O}'$ which the  3-vector velocity $\bm v$ relative to  ${\cal O}$
is perpendicular  to the plane surface that contains the 3-vectors electric and magnetic induction, we must have
 $ {\bm v}\cdot {\bm E} = {\bm v}\cdot {\bm B} = 0$. Also, we can use the common and alternative parametrization:
\begin{eqnarray}
\label{velocidade_de_O'}
 \gamma &=& \cosh \phi\cr
 \bm v \gamma &=& \hat{\bm v} \sinh\phi \cr
 \bm v  &=& \hat{\bm v} \tanh\phi \hspace*{1cm} \mbox{where}\,\, \hat{\bm v}=\frac{\bm v}{|\bm v|},
\end{eqnarray}
to rewrite the Lorentz transformations (\ref{transformacoes_Lorentz_E_B}) as follow:
\begin{eqnarray}
\label{transformacoes_Lorentz_E_B_2}
 {\bm E}' &=& {\bm E} \cosh\phi + \hat{\bm v} \times {\bm B} \sinh\phi \cr
  i{\bm B}' &=& i{\bm B} \cosh\phi  - i\hat{\bm v}\times {\bm E}\sinh\phi. 
\end{eqnarray} 
From the definition (\ref{vetor_eletromagnetico_complexo_2}) we can obtain from above two equations the 
following equation:
\begin{equation}
\label{transformacoes_Lorentz_E_B_3}
  \mbox{\bf F}' =  \mbox{\bf F} \cosh\phi - i \hat{\bm v}\times \mbox{\bf F} \sinh\phi .
\end{equation}

We have seen that the magnitude of a 4-vector is an invariant under Lorentz transformation (\ref{invariacia_1}),
\begin{equation}
\label{vetor_eletromagnetico_complexo_3}
 \mbox{F}_{\mu}\mbox{F}^{\mu} = 0 - \mbox{\bf F} \cdot \mbox{\bf F}
 =- \mbox{\bf F} \cdot \mbox{\bf F}.
\end{equation}
It is straightforward to show that  $ \mbox{\bf F}' \cdot \mbox{\bf F}' = \mbox{\bf F} \cdot \mbox{\bf F}$,
where we have
\begin{equation}
\label{identidade_4}
-|{\bm E}'|^2 + |{\bm B}'|^2 - i 2{\bm E}'\cdot {\bm B}'= -|{\bm E}|^2 + |{\bm B}|^2 - i 2{\bm E}\cdot {\bm B}.
\end{equation}
It is a complex relation where a real part $-|{\bm E}|^2 + |{\bm B}|^2= I_1$ is the invariant under 
Lorentz transformation which was seen in (\ref{invariante_1}). 
And similarly the imaginary part  $-2{\bm E}\cdot {\bm B}=I_2$ 
also is invariant under Lorentz transformation \cite{Landau}, in agreement with (\ref{invariante_2}).

From the invariant relations (\ref{vetor_eletromagnetico_complexo_3}) and (\ref{identidade_4})
we can state that:
\begin{enumerate}

 \item  if in an inertial frame $\cal O$ we have ${\bm E} \perp {\bm B}$, where the imaginary part 
 of $\mbox{\bf F} \cdot \mbox{\bf F}$ is vanished,
 then in any  inertial frame ${\cal O}'$ the electric field and magnetic induction field will 
 be also perpendicular vectors, ${\bm E}' \perp {\bm B}'$;
 
  \item  if in an inertial frame $\cal O$ we have $|{\bm E}| = |{\bm B}|$, where the real part 
 of $\mbox{\bf F} \cdot \mbox{\bf F}$ is vanished,
 then in any  inertial frame ${\cal O}'$ the magnitude of the electric field is equal to magnitude of the
 magnetic induction field, $|{\bm E}'| = |{\bm B}'|$;
 
  \item  if in an inertial frame $\cal O$ we have $|{\bm E}| < |{\bm B}|$ (or $|{\bm B}| < |{\bm E}|$), 
 then in any  inertial frame ${\cal O}'$ we have also  $|{\bm E}'| < |{\bm B}'|$ (or $|{\bm B}'| < |{\bm E}'|$);
 
  \item  if in an inertial frame $\cal O$ the angle between ${\bm E}$ and ${\bm B}$ is acute angle
(or obtuse angle), then in any  inertial frame ${\cal O}'$ the angle between  electric field ${\bm E}'$ 
and  magnetic induction field  ${\bm B}'$ is also acute angle (or obtuse angle);

  \item  if in an inertial frame $\cal O$ we have ${\bm E} = 0$ (or ${\bm B} = 0$), 
 then in any  inertial frame ${\cal O}'$ we have  ${\bm E}' \perp {\bm B}'$

  \item  if in an inertial frame $\cal O$ the imaginary part 
 of $\mbox{\bf F} \cdot \mbox{\bf F}$ is vanished, ${\bm E}\cdot {\bm B} = 0$ such that ${\bm E} =0$ 
 but ${\bm B} \neq 0$  the electromagnetic field is purely magnetic, 
 or ${\bm B} =0$ but ${\bm E} \neq 0$, the electromagnet field is purely electric, 
 then in any  inertial frame ${\cal O}'$ we have $|{\bm E}'|^2 - |{\bm B}'|^2 <0$, 
 the electromagnetic field is purely magnetic, or  $|{\bm E}'|^2 - |{\bm B}'|^2 >0$, 
 the electromagnetic field is purely electric;
 
   \item if in an inertial frame $\cal O$ we have the invariant $\mbox{\bf F} \cdot \mbox{\bf F} = 0$,
where $-|{\bm E}|^2 + |{\bm B}|^2 =0$ and ${\bm E}\cdot {\bm B} = 0$ , the electromagnetic field is said
to be null or singular and consequently the electromagnetic field will be null in any inertial frame 
 ${\cal O}'$.
 
\end{enumerate}

Now there is an important result associated with vector product between complex vector ${\bm F}$ with its 
complex conjugate ${\bm F}*$. From equation  (\ref{vetor_eletromagnetico_complexo_2}), we have
 \begin{equation}
 \label{F x F*}
   \mbox{\bf F} \times  \mbox{\bf F}^{*} = ({\bm E} + i{\bm B})\times ({\bm E} - i{\bm B})
   = - i {\bm E} \times {\bm B } +i{\bm B}\times {\bm E} = -i2 {\bm E} \times {\bm B }.
 \end{equation}
The relation between two inertial frames results in
\begin{equation}
  \mbox{\bf F}' \times \mbox{\bf F}'^{*} =  (\mbox{\bf F} \cosh\phi - i \hat{\bm v}\times \mbox{\bf F} \sinh\phi)
  \times  (\mbox{\bf F}^{*} \cosh\phi + i \hat{\bm v}\times \mbox{\bf F}^{*} \sinh\phi) \nonumber
\end{equation}
or
\begin{equation}
 -i2 {\bm E}' \times {\bm B }' = \mbox{\bf F}\times\mbox{\bf F}^{*} \cosh^2\phi +
 [ \mbox{\bf F} \times (\hat{\bm v}\times \mbox{\bf F}^{*}) -  (\hat{\bm v}\times \mbox{\bf F})\times \mbox{\bf F}^{*}
 ]i\sinh\phi\,\cosh\phi 
 +  (\hat{\bm v}\times \mbox{\bf F})\times (\hat{\bm v}\times \mbox{\bf F}^{*}) \sinh^2\phi.\nonumber
\end{equation}
\newpage
\noindent
Again, if we choose an inertial system ${\cal O}$ which the  3-vector velocity $\bm v$ relative to  ${\cal O}'$
is perpendicular  to the plane surface that contains the 3-vectors electric and magnetic induction, we must have
$\hat{\bm v} \perp \mbox{\bf F}$ in inertial frame ${\cal O}$ such as
\begin{equation}
 -i2 {\bm E}' \times {\bm B }' = \mbox{\bf F} \times  \mbox{\bf F}^{*} (\cosh^2 \phi +\sinh^2\phi)
 +i2\sinh\phi\cosh\phi (\mbox{\bf F} \cdot  \mbox{\bf F}^{*})\hat{\bm v},\nonumber
\end{equation}
with $\mbox{\bf F} \cdot  \mbox{\bf F}^{*} = |{\bm E}|^2+ |{\bm B}|^2$, it results in
\begin{equation}
 -i2 {\bm E}' \times {\bm B }' = -i2 {\bm E} \times {\bm B } \cosh(2 \phi)
 +i (|{\bm E}|^2+ |{\bm B}|^2)\hat{\bm v} \, \sinh(2\phi),\nonumber
\end{equation}
or
\begin{equation}
\label{E x B}
 2 {\bm E}' \times {\bm B }' = 2 {\bm E} \times {\bm B } \cosh(2 \phi)
 - (|{\bm E}|^2+ |{\bm B}|^2)\hat{\bm v} \, \sinh(2\phi).
\end{equation}
Let us consider an inertial frame  ${\cal O}'$, where the complex vector of electrodynamic field {\bf F} and 
a unit complex vector $\hat{\bm n}$ are parallel to each other, i.e.,
\begin{equation}
\label{E_paralelo_B_1}
 \mbox{\bf F}' = ( |{\bm E}'| + i  |{\bm B}'| ) \hat{\bm n},
\end{equation}
hence  $\hat{\bm n}$ is given by:
\begin{equation}
 \hat{\bm n} = \cosh\vartheta\,\, \hat{\bm e}_1 +i\sinh\vartheta\,\, \hat{\bm e}_2. 
\end{equation}
Then, the Lorentz invariant $ \mbox{\bf F}' \cdot  \mbox{\bf F}'$ is:
\begin{equation}
\label{E_paralelo_B_2}
 \mbox{\bf F}'  \cdot  \mbox{\bf F}' = |{\bm E}'|^2 - |{\bm B}'|^2 + 2i|{\bm E}'||{\bm B}'|,
\end{equation}
The imaginary part of above equation ($ 2|{\bm E}'||{\bm B}'| = 2{\bm E}'\cdot{\bm B}'$) shows that 
$ {\bm E}' \times {\bm B }' =0$. Thus, we can choose a Lorentz frame of reference ${\cal O}'$ in which 
$ {\bm E}'$ and ${\bm B }'$ are parallel and therefore we obtain for other Lorentz frame of reference ${\cal O}$
in which from (\ref{E x B}) we find that:
\begin{equation}
\label{velocidade_do_campo}
 \frac{2 {\bm E} \times {\bm B } }{|{\bm E}|^2+ |{\bm B}|^2} = \tanh(2\phi)\hat{\bm v}.
\end{equation}
It is worth noting that if we use Lorentz transformations (\ref{transformacoes_Lorentz_E_B}), where
${\bm E}'$ is parallel to ${\bm B}'$, ${\bm v}\perp {\bm E}'$ and ${\bm v}\perp {\bm B}'$ we have:
\begin{equation}
 {\bm E}'\times {\bm B}'= \gamma^2\left( {\bm E}+ {\bm v}\times {\bm B}\right)
 \times \left( {\bm B} - {\bm v}\times {\bm E}\right)=0 \nonumber
\end{equation}
that results in:
\begin{equation}
 \frac{ {\bm E} \times {\bm B } }{|{\bm E}|^2+ |{\bm B}|^2} = \frac{\bm v}{1+v^2}.
\end{equation}
In this Lorentz frame $\cal O$ we have that 
\begin{equation}
 \tanh(2\phi) = \frac{2\bm v}{1+v^2}.\nonumber
\end{equation}

For the electromagnetic plane wave in vacuum, the ratio between Poynting vector ${\bm S}$ and density of 
energy $u$ results in:
\begin{equation}
 \frac{2 {\bm E} \times {\bm B } }{|{\bm E}|^2+ |{\bm B}|^2} = c\,{\hat{\bm v}},
\end{equation}
while for a superposition of plane waves going in different directions 
the ratio is less than speed of the light.

The same field examined by an observer in  Lorentz frame  ${\cal O}'$ moving with velocity 
${\bm v}$ with respect to the frame of reference ${\cal O}$, previously used, being ${\bm v}$
as established in equation (\ref{velocidade_de_O'}):
\begin{equation}
{\bm v}=\tanh\phi\,\,\hat{\bm v},\nonumber
\end{equation}
perpendicular to the direction of both vectors: electric ${\bm E}$ and magnetic ${\bm B}$ in the frame 
${\cal O}$, in the Lorentz frame ${\cal O}'$ the Poynting flux  is zero. The electric ${\bm E}'$
and magnetic ${\bm B}'$ must be parallel. 

\subsection{Duality Rotation}

Let ${\bm f}$ be a complex electromagnetic vector defined by
\begin{equation}
 {\bm f} ={\bm e} +i {\bm b}.
\end{equation}
We can consider that this complex electromagnetic vector is a result from a complex rotation
of vector $\mbox{\bf F}={\bm E}+ i{\bm B}$,
\begin{equation}
 \label{rotacao_dual_1}
 {\bm f} = \mbox{\bf F}\,\, e^{-i\alpha},
\end{equation}
such that,
\begin{equation}
 {\bm e} +i{\bm b} = ({\bm E}\cos\alpha +{\bm B}\sin\alpha) + i( -{\bm E}\sin\alpha +{\bm B}\cos\alpha)
\end{equation}
so that,
\begin{eqnarray}
\label{rotacao_dual_2}
 {\bm e} &=& {\bm E}\cos\alpha +{\bm B}\sin\alpha \cr
 {\bm b} &=& -{\bm E}\sin\alpha +{\bm B}\cos\alpha.
\end{eqnarray}
We can choose the condition where 
${\bm e}\perp  {\bm b}$, such that:
\begin{eqnarray}
 (-|{\bm E}|^2+ |{\bm B}|^2)\sin\alpha\cos\alpha + {\bm E}\cdot {\bm B}(\cos^2\alpha -\sin^2\alpha)&=& 0\cr
 (-|{\bm E}|^2+ |{\bm B}|^2)\frac{\sin(2\alpha)}{2} + {\bm E}\cdot {\bm B}\cos(2\alpha) &=& 0,\nonumber
\end{eqnarray}
it follows that:
\begin{equation}
\label{condicao_extrema_1}
 \tan(2\alpha)= \frac{2 {\bm E}\cdot {\bm B}}{|{\bm E}|^2 -|{\bm B}|^2 }.
\end{equation}
From (\ref{campo_eletrico_2}) and (\ref{campo_magnetico_2}), where 
$E_{\mu}=F_{\mu\nu}u^{\nu}$ e $B_{\mu} = \widetilde{F}_{\mu\nu} u^{\nu}$, 
we can arrive  at a conclusion that the complex electromagnetic vector
${\bm f}$  also is a result of a contraction of a 4-velocity $u^{\nu}$ with a self-dual bivector,
\begin{equation}
 {\bm f} \rightarrow f_{\mu}=f_{\mu\nu} u^{\nu} + i \widetilde{f}_{\mu\nu} u^{\nu} = e_{\mu}+
 ib_{\mu},
\end{equation}
such as
\begin{equation}
 {\bm e}\rightarrow e_{\mu} = f_{\mu\nu} u^{\nu} \hspace{1cm}\mbox{and}\hspace{1cm}
  {\bm b} \rightarrow b_{\mu} = \widetilde{f}_{\mu\nu} u^{\nu}.
\end{equation}
It follows from above assumptions and (\ref{rotacao_dual_1}) that we have the resulting transformation or 
{\it duality rotation}:
\begin{equation}
\label{rotacao_dual_2.1}
 f_{\mu\nu} + i\widetilde{f}_{\mu\nu} = \left(F_{\mu\nu} + i\widetilde{F}_{\mu\nu}\right) e^{-i\alpha} =
 {\cal F}_{\mu\nu} e^{-i\alpha}
\end{equation}
or 
\begin{eqnarray}
 \label{rotacao_dual_3}
 f_{\mu\nu} &=& F_{\mu\nu}\,\cos{\alpha} + \widetilde{F}_{\mu\nu}\,\sin\alpha \cr
 \widetilde{f}_{\mu\nu} &=& -F_{\mu\nu}\,\sin{\alpha} + \widetilde{F}_{\mu\nu}\,\cos\alpha
\end{eqnarray}
at each point of spacetime. 

If we evaluate the contraction $f_{\mu\nu} \widetilde{f}^{\mu\nu}$ as follows,
\begin{eqnarray}
 f_{\mu\nu} \widetilde{f}^{\mu\nu} &=& -F_{\mu\nu}F^{\mu\nu}\,\sin{\alpha} \,\cos{\alpha} +
 \widetilde{F}_{\mu\nu}\widetilde{F}^{\mu\nu}\,\sin\alpha \,\cos\alpha
 + F_{\mu\nu}\widetilde{F}^{\mu\nu}\,\cos^2{\alpha} -  \widetilde{F}_{\mu\nu}F^{\mu\nu}\,\sin^2\alpha \cr
  &=& -F_{\mu\nu}F^{\mu\nu}\sin{(2\alpha)} +   F_{\mu\nu}\widetilde{F}^{\mu\nu} \cos{(2\alpha)}\nonumber
\end{eqnarray}
with equations (\ref{invariante_1}) and (\ref{invariante_2}) results in
\begin{equation}
 f_{\mu\nu} \widetilde{f}^{\mu\nu}  = -2(|{\bm B}|^2 -  |{\bm E}|^2)\sin{(2\alpha)}
 - 4 {\bm E}\cdot {\bm B} \cos{(2\alpha)},\nonumber
\end{equation}
we can obtain (\ref{condicao_extrema_1}) if we impose the  below condition:
\begin{equation}
 \label{condicao_extrema_2}
 f_{\mu\nu} \widetilde{f}^{\mu\nu}  = 0 \hspace{2cm} \mbox{(extremal field)}.
\end{equation}
Then, if we consider an angle $\alpha$ in a duality rotation such  that the equations  (\ref{condicao_extrema_1}) 
and (\ref{condicao_extrema_2}) are satisfied, 
we have a coordinate system where ${\bm e}\perp  {\bm b}$. 
In accordance with item 5 of invariant relations (\ref{vetor_eletromagnetico_complexo_3}) and (\ref{identidade_4}),
if we have ${\bm e}\perp  {\bm b}$, then at any point in spacetime with Minkowski metric, 
we can make a Lorentz transformation such that for a general (non-null) field at that point, ${\bm b}'=0$
and $ {\bm e}'$ is parallel to the $x-$axis, therefore in this Lorentz system of coordinates we have
$f'_{\mu\nu}f'^{\mu\nu}=-|{\bm e}'|<0$, then the electromagnetic field is purely electric.

The inverse transformations are:
\begin{eqnarray}
\label{rotacao_dual_4}
  F_{\mu\nu} &=& f_{\mu\nu}\,\cos{\alpha} - \widetilde{f}_{\mu\nu}\,\sin\alpha \cr
  \widetilde{F}_{\mu\nu} &=& f_{\mu\nu}\,\sin{\alpha} + \widetilde{f}_{\mu\nu}\,\cos\alpha.
\end{eqnarray}
From these transformations with extremal field condition (\ref{condicao_extrema_2}), we can obtain:
\begin{equation}
 F_{\mu\nu}F^{\mu\nu} = f_{\mu\nu}f^{\mu\nu} \cos(2\alpha)\nonumber
\end{equation}
and
\begin{equation}
 \widetilde{F}_{\mu\nu}F^{\mu\nu} = f_{\mu\nu}f^{\mu\nu} \sin(2\alpha),\nonumber
\end{equation}
such as 
\begin{equation}
  (F_{\mu\nu}F^{\mu\nu})^2 + (\widetilde{F}_{\mu\nu}F^{\mu\nu})^2 = ( f_{\mu\nu}f^{\mu\nu})^2\nonumber
\end{equation}
and so, we get the following:
\begin{equation}
\label{condicao_extrema_3}
( f_{\mu\nu}f^{\mu\nu}) = 
-\sqrt{(F_{\mu\nu}F^{\mu\nu})^2 + (\widetilde{F}_{\mu\nu}F^{\mu\nu})^2}
\end{equation}
where $f_{\mu\nu}$ demands the minus sign because it is the pure electric field along the $x$-axis.

The above transformation from $f_{\mu\nu}$ to  $F_{\mu\nu}$ in accordance with  (\ref{rotacao_dual_1}),
was discussed by Rainich in 1925 and was called
a {\it duality rotation} by Misner and Wheeler in 1957. Misner and Wheeler called the field $f_{\mu\nu}$
with the properties $ f_{\mu\nu} \widetilde{f}^{\mu\nu}  = 0$ an {\it extremal field}. The actual 
electromagnetic field $F_{\mu\nu}$ at a point can be obtained from the extremal field by a duality rotation
through the angle $\alpha$, where $\alpha$ is called the `{\it complexion}' of electromagnetic field at the point.

\subsection{Classical Field Theory of Electromagnetism}
In Classical Field Theory it is convenient and useful for many purpose to have Lagrangian formulation
of Field Theory which plays such a central role in the contemporary understanding of interactions and 
symmetries. The notion of a Lagrangian formulation of Field Theory is closely analogous to that of a 
Lagrangian formulation in Classical Mechanics. The action $S$ of any field on Classical Field Theory
is given by 
\begin{equation}
 \label{action_field_1}
 S= \int d^4 x {\cal L},
\end{equation}
where ${\cal L}$ is called a Lagrangian density.

The electromagnetic field equations in Minkowski spacetime (\ref{Maxwell_equations_1}) are derivable from a Lagrangian, 
which may be expressed in terms of the Lorentz covariant vector potential $A^{\mu}$:
\begin{equation}
 {\cal L}=-\frac{1}{4}F_{\rho\sigma}F^{\rho\sigma} -j^{\rho}A_{\rho}.
\end{equation}
The principle of least action states that the equation of motion is the one for which action $S$ is a
minimum, yields 
then the  Euler-Lagrange equations
\begin{equation}
 \frac{\partial {\cal L}}{\partial A_{\mu}}-  \partial_{\rho}
 \left(\frac{\partial {\cal L}}{\partial(\partial_{\rho} A_{\mu})}\right) = 0.
\end{equation}
It is a way of deriving inhomogeneous Maxwell equations. 
For each component $A_{\mu}$ give 
$$ \partial_{\rho}F^{\rho\mu}= -j^{\mu}.$$


The {\it Noether Theorem} states that for every symmetry of the action of a field there exist a corresponding
conserved quantity. In particular, if an action is invariant under a spacetime translation, characterized by 
a coordinate transformation of the form $x^{\mu} \rightarrow x^{\mu} + a^{\mu}$ in which
the vector $a^{\mu}$ does not depend on spacetime position, then one can define the 
 `canonical Noether energy-momentum tensor'.
Using the Lagrangian formulation of field
theory, the canonical Noether energy-momentum tensor for Lagrangian of electromagnetic field is found to be
\begin{equation}
\label{Noether_Tensor}
 {t^{\mu}}_{\nu}=  \left(-\frac{\partial {\cal L}}{\partial(\partial_{\mu} A_{\rho})}\right) \partial_{\nu}A_{\rho}
 +{\delta^{\mu}}_{\nu}{\cal L}.
\end{equation}
The Noether theorem yields the energy-momentum tensor for electromagnetic field free of sources, 
$j^{\mu}=0$,
\begin{equation}
 t_{\mu\nu}= F_{\mu\rho}\partial_{\nu}A^{\rho}-\frac{1}{4} \eta_{\mu\nu} F_{\rho\sigma}F^{\rho\sigma}.
\end{equation}
However, the canonical Noether energy-momentum tensor of the electromagnetic field is
asymmetric and not $U(1)$-gauge invariant and one has to fix it by means of the 
Belinfante-Rosenfeld procedure \cite{Blagojevic,Weinberg}:
\begin{equation}
\label{tensor_Belinfante_1}
T_{\mu\nu} = t_{\mu\nu} -\frac{i}{2}\partial^{\lambda}\left(S_{\mu\nu\lambda} + S_{\lambda\mu\nu}
- S_{\nu\lambda\mu}\right),
\end{equation}
where $S_{\mu\nu\lambda} $ is the contribution of the intrinsic (spin) angular momentum tensor, given by:
\begin{equation}
 S_{\mu\nu\lambda} = \frac{\partial{\cal L}}{\partial(\partial^{\mu}A^{\rho})}
 {\left[\Sigma_{\nu\lambda}\right]^{\rho}}_{\sigma}A^{\sigma},
\end{equation}
and  ${\left[\Sigma_{\nu\lambda}\right]^{\rho}}_{\sigma}$ are Hermitian generators of the group $SO(1,3)$ of
Lorentz transformation (\ref{gerador_lorentz}) applied to  vector potential  $A^{\mu}$ massless and spin 1.

Then, we can calculate $S_{\mu\nu\lambda} $, for Lagrangian of electromagnetic field free of sources, 
${\cal L}=-\dfrac{1}{4}F_{\tau\sigma}F^{\tau\sigma}$,
\begin{eqnarray}
 S_{\mu\nu\lambda} &=& -\frac{1}{2}F_{\tau\sigma}\left({\delta_{\mu}}^{\tau}{\delta_{\rho}}^{\sigma} -
 {\delta_{\mu}}^{\sigma}{\delta_{\rho}}^{\tau}\right){\left[\Sigma_{\nu\lambda}\right]^{\rho}}_{\sigma}A^{\sigma}\cr
  &=& -\frac{1}{2}(F_{\mu\rho}-F_{\rho\mu})i\left({\delta_{\lambda}}^{\rho}\eta_{\nu\sigma} - 
{\delta_{\nu}}^{\rho}\eta_{\lambda\sigma}\right)A^{\sigma}\cr
&=& -iF_{\mu\rho}\left({\delta_{\lambda}}^{\rho}A_{\nu} - 
{\delta_{\nu}}^{\rho}A_{\lambda}\right)\cr
&=& -i\left(F_{\mu\lambda}A_{\nu} - F_{\mu\nu}A_{\lambda}\right).\nonumber
\end{eqnarray}
Therefore, 
\begin{equation}
 S_{\mu\nu\lambda} + S_{\lambda\mu\nu}
- S_{\nu\lambda\mu} = -2i  F_{\mu\lambda}A_{\nu}.\nonumber
\end{equation}
For electromagnetic field free of sources, $\partial^{\lambda}F_{\mu\lambda}=0$, we obtain,
\begin{equation}
 \partial^{\lambda}\left(S_{\mu\nu\lambda} + S_{\lambda\mu\nu}
- S_{\nu\lambda\mu}\right) =-2iF_{\mu\lambda}\partial^{\lambda}A_{\nu}.\nonumber
\end{equation}
We can put the above result in the equation of Belinfante-Rosenfeld tensor (\ref{tensor_Belinfante_1}) such as:
\begin{equation}
T_{\mu\nu} = \, t_{\mu\nu} -\frac{i}{2}\left(-2iF_{\mu\lambda}\partial^{\lambda}A_{\nu}\right),\nonumber
\end{equation}
that results in,
\begin{eqnarray}
T_{\mu\nu} &=&  F_{\mu\rho}\partial_{\nu}A^{\rho}-\frac{1}{4} \eta_{\mu\nu} F_{\rho\sigma}F^{\rho\sigma}
-F_{\mu\rho}\partial^{\,\rho}A_{\nu}\cr
&=&  F_{\mu\rho}\left(\partial_{\nu}A^{\rho}- \partial^{\,\rho}A_{\nu}\right)-
\frac{1}{4} \eta_{\mu\nu} F_{\rho\sigma}F^{\rho\sigma}, \nonumber
\end{eqnarray}
now,  the energy-momentum tensor is symmetric and  $U(1)$-gauge invariant:
\begin{equation}
\label{tensor_EM_F_1}
 T_{\mu\nu}= F_{\mu\rho}{F_{\nu}}^{\rho} - \frac{1}{4} \eta_{\mu\nu} F_{\rho\sigma}F^{\rho\sigma}.
\end{equation}
In Special Relativity and General Relativity, the fields are described by energy-momentum tensor
$T_{\mu\nu}$, and it satisfies  the conservation law:
\begin{equation}
 \partial^{\mu}T_{\mu\nu} =0,
\end{equation}
it means that energy and momentum are conserved quantities.

We can arrange the components of energy-momentum tensor of electromagnetism into a symmetric matrix.
We can calculate the first term of $T_{\mu\nu}$ from above equation,  $  F_{\mu\rho}{F_{\nu}}^{\rho}$,
where we can use   $  F_{\mu\rho}{F_{\nu}}^{\rho} = -  F_{\mu\rho}{F^{\rho}}_{\nu}$ to facilitate our
matrix calculations:
\begin{eqnarray}
 ( F_{\mu\rho}{F^{\rho}}_{\nu})= 
 \begin{pmatrix}
  0 & -E_x & -E_y & -E_z \cr
   E_x & 0 & B_z & -B_y \cr
   E_y & -B_z & 0 & B_x \cr
   E_z & B_y & -B_x & 0 \cr
 \end{pmatrix} \cdot
 \begin{pmatrix}
  0 & E_x & E_y & E_z \cr
   E_x & 0 & B_z & -B_y \cr
   E_y & -B_z & 0 & B_x \cr
   E_z & B_y & -B_x & 0 \cr
 \end{pmatrix},\nonumber
\end{eqnarray}
it reduces to 
\begin{eqnarray}
  (F_{\mu\rho}{F_{\nu}}^{\rho}) = -( F_{\mu\rho}{F^{\rho}}_{\nu})= -
 \begin{pmatrix}
  -|{\bm E}|^2 & B_zE_y-B_yE_z & B_xE_z-B_z E_x & B_yE_x-B_xE_y \cr 
  & & & \cr
  B_zE_y-B_yE_z & E_x^2 -B_ y^2- B_z^2 & E_xE_y+B_xB_y & E_xE_z+ B_xB_z \cr
  & & & \cr
   B_xE_z-B_z E_x & E_xE_y+B_xB_y & E_y^2-B_x^2 -B_z^2 & E_yE_z+B_yB_z\cr
   & & & \cr
   B_yE_x-B_xE_y &  E_xE_z+ B_xB_z & E_yE_z+B_yB_z & E_z^2 - B_x^2-B_y^2
 \end{pmatrix}.\nonumber
\end{eqnarray}
Now, recall that $ \frac{1}{4} \eta_{\mu\nu} F_{\rho\sigma}F^{\rho\sigma}= 
 \frac{1}{2} \eta_{\mu\nu}(|{\bm B}|^2 -|{\bm E}|^2)$. And so, from (\ref{tensor_EM_F_1}) it leads to matrix
 of energy-momentum tensor:
\begin{equation}
(T_{\mu\nu}) = 
 \begin{pmatrix}
  \frac{1}{2}(|{\bm E}|^2 +|{\bm B}|^2) & -({\bm E}\times{\bm B})_x & -({\bm E}\times{\bm B})_y & 
  -({\bm E} \times {\bm B})_z \cr 
  & & & \cr
 -({\bm E}\times{\bm B})_x & \frac{1}{2}(|{\bm E}|^2+|{\bm B}|^2)-E_x^2 -B_x^2  & 
 -E_xE_y-B_xB_y & -E_xE_z- B_xB_z \cr
  & & & \cr
   -({\bm E}\times{\bm B})_y & -E_xE_y-B_xB_y & 
   \frac{1}{2}(|{\bm E}|^2+|{\bm B}|^2)-E_y^2 -B_y^2 & - E_yE_z-B_yB_z\cr
   & & & \cr
   -({\bm E} \times {\bm B})_z &  -E_xE_z - B_xB_z & -E_yE_z-B_yB_z 
   &  \frac{1}{2}(|{\bm E}|^2+|{\bm B}|^2)-E_z^2 -B_z^2 
 \end{pmatrix}.\nonumber
\end{equation} 
Let us describe each type of component from $T_{\mu\nu}$
\begin{itemize}
 \item  the time-time component represents energy density, $T_{00} = \frac{1}{2} (|{\bm E}|^2 +|{\bm B}|^2)$;
 
 \item the time-space components represent the energy flow. For example: 
 $T_{01} = -({\bm E}\times{\bm B})_x$ is $x$ component of Poynting vector;
 
 \item the space-time components represent the momentum density 
 (that is proportional to the Poynting flux \cite{Jackson}).
For example: 
 $T_{10} = -({\bm E}\times{\bm B})_x$ is the momentum density in the $x$ direction;
 
\item the spatial components represent stress. For example, 
$T_{11} =\sigma_{xx}$ is the force exerted in $x$ direction, per unit area normal to $x$, by electromagnetic field,
it is pressure. The component $T_{12} =\sigma_{xy}= -E_xE_y-B_xB_y$ is the force exerted in $x$ direction,
per unit area normal to $y$, by electromagnetic field, it is shear.
\end{itemize}
The purely spatial components $T_{ij}$ of energy-momentum of electromagnetic field is the definition
of the Maxwell stress tensor $\sigma_{ij}$ defined by \cite{Jackson}:
\begin{equation}
 \sigma_{ij}=\frac{1}{2}\delta_{ij}\left( |{\bm E}|^2+|{\bm B}|^2  \right)-E_i E_j - B_iB_j.
\end{equation}
We can simplify the energy-momentum of electromagnetic field as:
\begin{equation}
 (T_{\mu\nu})=
 \begin{pmatrix}
\label{tensor_energia_momento_eletromagnetico_3}
  \frac{1}{2}(|{\bm E}|^2 +|{\bm B}|^2) & -({\bm E}\times{\bm B})\cr
  -({\bm E}\times{\bm B}) & \sigma_{ij}
 \end{pmatrix}.
\end{equation}

Under Lorentz transformations (\ref{transformacoes_Lorentz_E_B}), the complex electromagnetic vector
 $\mbox{\bf F} = {\bm E} + i{\bm B}$ transforms in accordance with (\ref{transformacoes_Lorentz_E_B_3}),
since we can choose a Lorentz frame of reference where we have
  $ {\bm v}\perp {\bm E}$ and ${\bm v}\perp {\bm B} $. Then, if we choose the complex electromagnetic vector
from (\ref{E_paralelo_B_1}), such as,
\begin{equation}
 \mbox{\bf F}' = ( |{\bm E}'| + i  |{\bm B}'| ) \hat{\bm x}\hspace*{2cm} 
 \mbox{ in an inertial frame } {\cal O}',\nonumber
\end{equation} 
where we have  ${\bm E}'\parallel{\bm B}'$ or in other words we have: 
$${\bm E}' = E'_x\hat{\bm x}\hspace*{2cm} \mbox{and} \hspace*{2cm} {\bm B}' = B'_x\hat{\bm x},$$ 
therefore, we have $E'_y=E'_z=B'_y=B'_z=0$.
Now, in this inertial frame $ {\cal O}'$ we obtain the below matrix of energy-momentum tensor 
of electromagnetic field:
\begin{equation}
(T'_{\mu\nu}) = 
 \begin{pmatrix}
  \frac{1}{2}(|{\bm E}'|^2 +|{\bm B}'|^2) & 0& 0 & 0 \cr 
  & & & \cr
 0 & \frac{1}{2}(|{\bm E}'|^2+|{\bm B}'|^2)-{E'_x}^2 -{B'_x}^2  & 
0 & 0 \cr
  & & & \cr
  0 & 0 &  \frac{1}{2}(|{\bm E}'|^2+|{\bm B}'|^2) & 0\cr
   & & & \cr
  0 &  0 & 0
   &  \frac{1}{2}(|{\bm E}'|^2+|{\bm B}'|^2)
 \end{pmatrix}\nonumber
\end{equation}
or more explicitly,
\begin{equation}
\label{tensor_EM_F_3}
({T'_{\mu}}^{\nu}) = \frac{1}{2}(|{\bm E}'|^2 +|{\bm B}'|^2)
 \begin{pmatrix}
  -1 & 0& 0 & 0 \cr 
   0 & -1 & 0 & 0 \cr
   0 & 0 & 1 & 0 \cr
   0 &  0 & 0 &  1
 \end{pmatrix}.
\end{equation}
This energy-momentum tensor has two important properties: (i) its trace is zero and (ii) its square
is multiple of the unit matrix.
In this case, it is convenient to see the square ${T'_{\mu\rho}}{T'^{\rho\nu}}$:
\begin{equation}
 \label{quadrado_de_T}
{T'_{\mu\rho}}{T'^{\rho\nu}} = \frac{1}{4} \left(|{\bm E}'|^2 +|{\bm B}'|^2\right)^2{\delta_{\mu}}^{\nu},
\end{equation}
therefore, we can write:
$$  \left(|{\bm E}'|^2 +|{\bm B}'|^2\right)^2 =  \left(|{\bm E}'|^2 -|{\bm B}'|^2\right)^2 
+ 4 \left(|{\bm E}'||{\bm B}'|\right)^2.$$
We have affirmed that  ${\bm E}'\parallel{\bm B}'$, such as $|{\bm E}'||{\bm B}'| = {\bm E}'\cdot {\bm B}'$.
Thus, the above equation results in 
$$   \left(|{\bm E}'|^2 +|{\bm B}'|^2\right)^2 =   \left(|{\bm B}'|^2 - |{\bm E}'|^2\right)^2 
+  (2{\bm E}'\cdot{\bm B}')^2.$$
We recall the Lorentz invariants from (\ref{invariante_1}) and (\ref{invariante_2}), where we have:
\begin{equation}
 I_1 = \frac{1}{2}F\,'_{\mu\nu}{F\,'}^{\mu\nu}= |{\bm B}'|^2 - |{\bm E}'|^2 \hspace*{1cm}
 \mbox{and} \hspace*{1cm} 
 I_2 = \frac{1}{2}F\,'_{\mu\nu}\tilde{F\,'}^{\mu\nu}= -2{\bm E}'\cdot{\bm B}'.\nonumber
\end{equation}
Now, we obtain that
\begin{equation}
\label{invariante_3}
  \left(|{\bm E}'|^2 +|{\bm B}'|^2\right)^2 =  I_1^2+ I_2^2.
\end{equation}
Finally, we can rewrite the square of energy-momentum tensor in an Lorentz reference system ${\cal O}'$ as
\begin{equation}
 \label{quadrado_de_T_2}
{T'_{\mu\rho}}{T'^{\rho\nu}} = \frac{1}{4} \left(I_1^2+ I_2^2\right)\,\,{\delta_{\mu}}^{\nu}.
\end{equation}
The above equation says that the square of the energy-momentum tensor is invariant to change of coordinate system.
It hold whether the energy-momentum tensor of electromagnetic field is diagonal or not.

Again, if both electromagnetic invariants $ I_1 $ and $ I_2 $ vanish, where we have $|{\bm E}|=|{\bm B}|$ and 
${\bm E}\perp {\bm B}$ respectively, the electromagnetic field is said to be null (or singular). For the situation where
$ I_1^2+ I_2^2 \neq 0$, the electromagnetic field is called the general case.

From equation (\ref{quadrado_de_T_2}), we have that:
\begin{equation}
{T'_{\mu\rho}}{T'^{\mu\rho}} = I_1^2+ I_2^2,\nonumber
\end{equation}
such that the equation (\ref{quadrado_de_T_2}) can be rewritten as:
\begin{equation}
 \label{quadrado_de_T_3}
{T'_{\mu\rho}}{T'^{\rho\nu}} = \frac{1}{4} \,\,{\delta_{\mu}}^{\nu} \left(
 {T'_{\kappa\lambda}}{T'^{\kappa\lambda}}\right).
\end{equation}
With (\ref{invariante_3}) it is possible to find a Minkowski system frame, in the general case, that puts
the tensor ${T_{\mu}}^{\nu}$ from (\ref{tensor_EM_F_3}) into the diagonal form:
\begin{equation}
\label{tensor_EM_F_4}
({T_{\mu}}^{\nu}) = \frac{1}{2}\sqrt{I_1^2+ I_2^2}\,\,
 \begin{pmatrix}
  -1 & 0& 0 & 0 \cr 
   0 & -1 & 0 & 0 \cr
   0 & 0 & 1 & 0 \cr
   0 &  0 & 0 &  1
 \end{pmatrix}.
\end{equation}
We choose the positive root of $|{\bm E}'|^2 +|{\bm B}'|^2$ that $T_{\mu\nu}u^{\mu}u^{\nu} \geq 0$
 be obeyed (where $u_{\mu}u^{\mu}=-1$).

%

\section{Electromagnetic Field in Curved Spacetime}

In the words of the Robert M.Wald \cite{Wald}: 
the laws of Physics in General Relativity are governed by two basic principles: 
(1) the principle of general covariance,
which states that the metric, $g_{\mu\nu}$, and quantities derivable from it are the only spacetime quantities
that can appear in the equations of physics; (2) the requirement that the equations must reduce to the equations
satisfied in Special Relativity in the case where $g_{\mu\nu} \rightarrow \eta_{\mu\nu}$ is flat.

Since one can identify spacetime curvature with gravitation, one can link a physical interaction
with gravitation by replace everywhere the metric $\eta_{\mu\nu}$ of Special Relativity by $g_{\mu\nu}$ of 
curved spacetime. Moreover, change all derivatives from ordinary derivatives to covariant ones.
It is a version of principle of equivalence that tells us how to write the equations of any physical
field interaction in a generally covariant form. One can use Special Relativity to write down the field 
equations in the locally inertial frame and change $\partial_{\mu}$ by $\nabla_{\mu}$,
allowing to write the field equations in curved spacetime. Such a link of a physical interaction 
with gravitation is called minimal coupling. We are now able to use it to rewrite the electromagnetic field
 (\ref{electromagnetic_tensor_1})
in the covariant language of curved spacetime:
\begin{equation}
 \label{electromagnetic_tensor_2}
 F_{\mu\nu}= \nabla_{\mu}A_{\nu} - \nabla_{\nu}A_{\mu}.
\end{equation}
Because of symmetries of the Christoffel symbols, (\ref{electromagnetic_tensor_2}) reduces 
(\ref{electromagnetic_tensor_1}), however.

The Maxwell source equations (\ref{Maxwell_equations_1}) change likewise:
\begin{equation}
\label{Maxwell_equations_2}
 \nabla_{\mu}F^{\mu\nu} = -j^{\nu}.
\end{equation}

In any arbitrary coordinates system, the invariant volume element is $dv=\sqrt{-g}d^4x$, where
$g$ is the determinant of the metric tensor $g_{\mu\nu}$ in that coordinate system. It is therefore
convenient to rewrite the action (\ref{action_field_1}) for electromagnetic field in curved space in the form
\begin{equation}
 \label{action_field_2}
 S= \int d^4 x \sqrt{-g} \left\{-\frac{1}{4}F_{\mu\nu}F_{\rho\sigma}g^{\mu\rho}g^{\nu\sigma} -j^{\rho}A_{\rho}
 \right\},
\end{equation}
or
\begin{equation}
 S= \int d^4 x {\cal L}, \nonumber
\end{equation}
where the field Lagrangian for electromagnetic is
\begin{equation}
 \label{Lagrangian_2}
 {\cal L} = \sqrt{-g} \left\{-\frac{1}{4}F_{\mu\nu}F_{\rho\sigma}g^{\mu\rho}g^{\nu\sigma} -
 j^{\rho}A_{\rho} \right\}.
\end{equation}
Demanding stationarity of the action, $\delta S=0$, one obtain the Euler-Lagrange equations for curved spacetime,
\begin{equation}
 \label{Euler_Lagrange_Equations_2}
  \frac{\partial {\cal L}}{\partial A_{\nu}}-  \nabla_{\mu}
 \left(\frac{\partial {\cal L}}{\partial(\nabla_{\mu} A_{\nu})}\right) = 0,
\end{equation}
that yields the covariant Maxwell source equation (\ref{Maxwell_equations_2}).

To bring the energy-momentum tensor of electromagnetic field which determines the curvature of spacetime, 
it is necessary to  regard
that the dynamical content of General Relativity is fully expressed by Einstein's field equation 
(\ref{field equation}). In a classical context, it is convenient and useful for many purposes to have
Lagrangian formulations of General Relativity. The Einstein-Hilbert action indeed yields the Einstein's
field equation under the variation with respect to the metric. The corresponding  Einstein-Hilbert action
for vacuum Einstein equation is
\begin{equation}
 \label{action_field_3}
 S_{EH} = \int d^4 x \sqrt{-g}\, R,
\end{equation}
where the Lagrangian of Einstein-Hilbert is given by
\begin{equation}
 \label{Lagrangian_EH}
 {\cal L}_{EH} = \sqrt{-g}\, R,
\end{equation}
and where $R$ is the trace of Ricci tensor. To obtain the coupled gravitational field with matter field,
one construct a total Lagrangian, by adding together the Einstein-Hilbert Lagrangian ${\cal L}_{EH}$
 with the Lagrangian ${\cal L}_{M}$, for the matter field,
${\cal L} = {\cal L}_{EH} + {\cal L}_{M}$, such that for action one have,
\begin{equation}
 \label{action_field_4}
 S = \int d^4 x \left\{ \frac{1}{2\kappa} {\cal L}_{EH} + {\cal L}_{M}\right\}.
\end{equation}
Since the Einstein-Hilbert Lagrangian ${\cal L}_{EH}$ does not depend on the matter field, variation of the total 
action $S$, with respect to metric $g_{\mu\nu}$ yields the equation,
\begin{equation}
\label{field equation_2}
R_{\mu\nu}-\frac{1}{2}g_{\mu\nu}R =\kappa T_{\mu\nu},
\end{equation}
where the tensor $T_{\mu\nu}$ is given by
\begin{equation}
 \label{Hilbert_energy-momentum_tensor}
 T_{\mu\nu} = \frac{2}{\sqrt{-g}}\frac{\delta{\cal L}_{M}}{\delta g^{\mu\nu}}.
\end{equation}
It is the `metric Hilbert energy-momentum' which is symmetric by definition and it is the quantity which
naturally appears on the right-hand side of Einstein's equation (\ref{field equation_2}) in a 
Lagrangian formulation of Einstein-matter field equations.

The  metric Hilbert energy-momentum for the source-free electromagnetic field such a Lagrangian
given by
$$ {\cal L}_{M} = - \frac{1}{4}F_{\kappa\lambda}F_{\rho\sigma}g^{\kappa\rho}g^{\lambda\sigma} \sqrt{-g}
= - \frac{1}{4}F^2\sqrt{-g},$$
where $F^2 = F_{\kappa\lambda}F_{\rho\sigma}g^{\kappa\rho}g^{\lambda\sigma}$, one find with following calculations
\begin{equation}
T^{\mu\nu} =
\frac{2}{\sqrt{-g}}\frac{\delta }{\delta g_{\mu\nu}}
\left(-\frac{1}{4} F_{\kappa\lambda}F_{\rho\sigma}g^{\kappa\rho}g^{\lambda\sigma}\sqrt{-g}\right),\nonumber
\end{equation}
$$
T^{\mu\nu} = \frac{2}{\sqrt{-g}}\frac{1}{\delta g_{\mu\nu}}
\left[ -\frac{1}{4}(2\delta g^{\kappa\rho})g^{\lambda\sigma} F_{\kappa\lambda} F_{\rho\sigma}\sqrt{-g}
-\frac{F^2}{4}\delta \sqrt{-g} \right],
$$
under the variation $ \delta \sqrt{-g} = -\frac{\sqrt{-g}}{2}g_{\kappa\lambda}\delta g^{\kappa\lambda}$,
one continue the calculations,
\begin{eqnarray}
T^{\mu\nu} &=& \frac{2}{\sqrt{-g}}\frac{1}{\delta g_{\mu\nu}}
\left[ -\frac{1}{2}{F_{\kappa}}^{\sigma} F_{\rho\sigma}\sqrt{-g}\,\,\delta g^{\kappa\rho}
-\frac{F^2}{4}\frac{\sqrt{-g}}{2}(-g_{\kappa\lambda}\delta g^{\kappa\lambda}) \right]\cr\cr
&=& \left[ -{F_{\kappa}}^{\sigma} F_{\rho\sigma}
+\frac{F^2}{4} g_{\kappa\rho} \right]\frac{\delta g^{\kappa\rho}}{\delta g_{\mu\nu}}\cr\cr
&=&  \left[ -{F_{\kappa}}^{\sigma} F_{\rho\sigma}
+\frac{F^2}{4} g_{\kappa\rho} \right](-g^{\kappa\mu}g^{\rho\nu})\cr\cr
&=& F^{\mu\sigma}{F^{\nu}}_{\sigma} - g^{\mu\nu} \frac{1}{4} F_{\rho\sigma}F^{\rho\sigma},\nonumber
\end{eqnarray}
it follows that in covariant form, the energy-momentum tensor for source-free electromagnetic field is
\begin{equation}
\label{tensor_EM_F_2}
T_{\mu\nu} =  F_{\mu\sigma}{F_{\nu}}^{\sigma} - g_{\mu\nu} \frac{1}{4} F_{\rho\sigma}F^{\rho\sigma}.
\end{equation}
In Classical Field Theory, for every symmetry of the action there exist a corresponding conserved quantity,
if an action is invariant under a spacetime translation, then one can define a  
canonical Noether energy-momentum tensor (\ref{Noether_Tensor}), that is neither gauge invariant nor
symmetric, but by the Belinfante-Rosenfeld procedure one can fix these drawbacks obtaining the 
symmetric energy-momentum tensor (\ref{tensor_EM_F_1}) in  Minkowski flat spacetime. Since the basic framework
of General Relativity modifies that of Special Relativity only in that it allows the manifold to differ from 
$\mathbb{R}^4$ and the metric to be nonflat, one may represent the energy-momentum tensor (\ref{tensor_EM_F_1}) 
of Special Relativity by metric Hilbert energy-momentum tensor of General Relativity (\ref{tensor_EM_F_2}), by
replace $\eta_{\mu\nu}$ by $g_{\mu\nu}$ and replace the derivative operator $\partial_{\mu}$ associated with
 $\eta_{\mu\nu}$ by the derivative operator $\nabla_{\mu}$ associated with $g_{\mu\nu}$.

 \subsection{Conserved Electric Current}
 
 The electric current $ j ^{\mu}$ is conserved in curved spacetime. This statement is verified by following
 commutator of the covariant derivative operators:
 \begin{equation}
  [\nabla_{\rho},\nabla_{\sigma} ]T_{\mu\nu} 
  = (\nabla_{\rho}\nabla_{\sigma} - \nabla_{\sigma}\nabla_{\rho})T_{\mu\nu}=
 {R_{\rho\sigma\mu}}^{\kappa}T_{\kappa\nu} + {R_{\rho\sigma\nu}}^{\kappa}T_{\mu\kappa},
 \end{equation}
that for the contraction with antisymmetric tensor of electromagnetism $F_{\mu\nu}$, we obtain
\begin{equation}
  (\nabla_{\mu}\nabla_{\nu} - \nabla_{\nu}\nabla_{\mu})F^{\mu\nu} =
 {R_{\mu\nu}}^{\mu\kappa}{F_{\kappa}}^{\nu} + {R_{\mu\nu}}^{\nu\kappa}{F^{\mu}}_{\kappa} =
 {R_{\mu\nu}}^{\mu\kappa}{F_{\kappa}}^{\nu} - {R_{\nu\mu}}^{\nu\kappa}{F^{\mu}}_{\kappa}  =
 {R_{\nu}}^{\kappa}{F_{\kappa}}^{\nu} - {R_{\mu}}^{\kappa}{F^{\mu}}_{\kappa} = -2R_{\mu\nu}F^{\mu\nu},
 \nonumber
\end{equation}
where the symmetric Ricci tensor contracted with the antisymmetric Maxwell tensor results in zero 
and consequently we have
\begin{equation}
  \nabla_{\mu}\nabla_{\nu}F^{\mu\nu} - \nabla_{\nu}\nabla_{\mu}F^{\mu\nu} = 0,\nonumber
\end{equation}
that with the Maxwell source equation (\ref{Maxwell_equations_2}), it results in
\begin{equation}
  \nabla_{\mu}j^{\mu} - \nabla_{\nu}(-j^{\nu}) = 2  \nabla_{\mu}j^{\mu} =0.\nonumber
\end{equation}
As shown in the above calculation, the Maxwell source equations in the presence of gravity, also have the
prescription of conserved electric charge $\nabla_{\mu}j^{\mu} =0$, just as they do in Special Relativity Theory.
 \subsection{Maxwell's Equations in the Lorentz Gauge}
In curved spacetime, the Maxwell's equations for vector potential $A_{\nu}$ in Lorentz gauge, are obtained
 as follows
\begin{equation}
 \nabla^{\mu}F_{\mu\nu}= \nabla^{\mu}(\nabla_{\mu}A_{\nu}-\nabla_{\nu}A_{\mu}) = -j_{\nu}, \nonumber
\end{equation}
or
\begin{equation}
\Box A_{\nu}-\nabla^{\mu}\nabla_{\nu}A_{\mu}=-j_{\nu}.\nonumber
\end{equation}
Hence, the covariant derivative are noncommutative operators, from identity:
\begin{equation}
\label{Maxwell_equations_3}
 (\nabla_{\mu}\nabla_{\nu} - \nabla_{\nu}\nabla_{\mu})A_{\rho} = R_{\mu\nu\rho\sigma}A^{\sigma},
\end{equation}
it is worth noting that if we contract above equation over $\mu$ and $\rho$, we find
\begin{equation}
 (\nabla^{\mu}\nabla_{\nu} - \nabla_{\nu}\nabla^{\mu})A_{\mu} = {R^{\mu}}_{\nu\mu\sigma}A^{\sigma}
 = R_{\nu\sigma}A^{\sigma},\nonumber
\end{equation}
which gives
\begin{equation}
 \nabla^{\mu}\nabla_{\nu}A_{\mu} =  \nabla_{\nu}\nabla^{\mu}A_{\mu} +  R_{\nu\sigma}A^{\sigma}.\nonumber
\end{equation}
Substituting above equation into Maxwell's equations (\ref{Maxwell_equations_3}), we obtain
\begin{equation}
\Box A_{\nu}- \nabla_{\nu}\nabla^{\mu}A_{\mu} -  R_{\nu\sigma}A^{\sigma}=-j_{\nu}. \nonumber
\end{equation}
A vector potential $A_{\nu}$ satisfying the Lorentz gauge condition $\nabla^{\mu}A_{\mu}=0$, reduces
the above equation to the final form
\begin{equation}
\label{Maxwell_equations_4}
\Box A_{\nu}-  R_{\nu\sigma}A^{\sigma} = -j_{\nu}.
\end{equation}
Above equation for vector potential $A_{\nu}$ contains an explicit term of curvature.

Although the electrodynamic equations are all obtained from Special Relativity by rule  
 $\eta_{\mu\nu} \rightarrow g_{\mu\nu}$ and $\partial_{\mu} \rightarrow \nabla_{\mu}$, 
the wave equation (\ref{Maxwell_equations_4}) for the vector potential is not. Nevertheless
when spacetime is flat, $R_{\nu\sigma}=0$, the wave equation  (\ref{Maxwell_equations_4}) does reduce
to the usual wave equation of Special Relativity.

\section{Einstein-Maxwell equations and the Rainich conditions}

In General Relativity, the energy-momentum tensor $T_{\mu\nu}$ acts as the source of the gravitational
field. It is related to the curvature of spacetime via the Einstein equation (\ref{field equation_2}).
When we have the field equations of General Relativity with energy-momentum tensor of the electromagnetic
field as the only source term, we can call this combination: `Einstein-Maxwell electrovac equations',
\begin{equation}
 \label{electrovac_1}
 R_{\mu\nu}-\frac{1}{2}g_{\mu\nu}R = \kappa\left(
 F_{\mu\sigma}{F_{\nu}}^{\sigma} - g_{\mu\nu} \frac{1}{4} F_{\rho\sigma}F^{\rho\sigma}\right).
\end{equation}
Note that the trace of energy-momentum tensor of the electromagnetic
field is null,
\begin{equation}
\label{traco_do_T_munu}
 T= {T_{\mu}} ^{\mu} =  F_{\mu\sigma}F^{\mu\sigma} - {\delta_{\mu}}^{\mu} \frac{1}{4} F_{\rho\sigma}F^{\rho\sigma}
  = F_{\mu\sigma}F^{\mu\sigma} - F_{\rho\sigma}F^{\rho\sigma} = 0.
\end{equation}

There is other way to write the electromagnetic energy-momentum tensor (\ref{tensor_EM_F_2}) in terms
of dual tensor of electromagnetic field $ \widetilde{F}_{\mu\nu}$.
From (\ref{identidade_2}), 
we have that
\begin{equation}
 F_{\mu\rho} {F_{\nu}}^{\rho} - \widetilde{F}_{\mu\rho}{\widetilde{F}_{\nu}}\,^{\rho} = 
 \frac{1}{2} g_{\mu\nu} F^{\sigma\tau}F_{\sigma\tau}, \nonumber
\end{equation}
that results in:
\begin{equation}
 \frac{1}{4}  g_{\mu\nu} F^{\sigma\tau}F_{\sigma\tau} = \frac{1}{2}\left( F_{\mu\rho} {F_{\nu}}^{\rho}
 -\widetilde{F}_{\mu\rho}{\widetilde{F}_{\nu}}\,^{\rho} \right).\nonumber
\end{equation}
Putting the above equation in the electromagnetic energy-momentum tensor (\ref{tensor_EM_F_2})
we have:
\begin{equation}
\label{tensor_energia_momento_eletromagnetico_2}
  T_{\mu\nu} = \frac{1}{2}\left( F_{\mu\rho} {F_{\nu}}^{\rho}
 + \widetilde{F}_{\mu\rho}{\widetilde{F}_{\nu}}\,^{\rho} \right).
\end{equation}
Now, let's recall the electromagnetic self-dual vector defined in (\ref{bivetor_complexo_3}),
where we have ${\cal F}_{\mu\nu}=F_{\mu\nu} + i \widetilde{F}_{\mu\nu}$,
and so, we get the following expression for electromagnetic energy-momentum tensor:
\begin{equation}
  T_{\mu\nu} = \frac{1}{2}\,{\cal F}_{\mu\rho}\left({{\cal F}_{\nu}}\,^{\rho}\right)^{*}.
\end{equation}

\subsection{The Rainich conditions}
When a curvature of spacetime is to be attributed to a source-free electromagnetic field,
from vanished trace of energy-momentum tensor (\ref{traco_do_T_munu}),  we have that
${R_{\mu}}^{\mu} -\frac{1}{2}{\delta_{\mu}}^{\mu}R = \kappa {T_{\mu}}^{\mu}=0$, such as:
\begin{equation}
 \label{condicao_Rainich_1}
R = 0.
\end{equation}
When this condition is satisfied, the Einstein curvature tensor on the left hand side of equation
(\ref{electrovac_1}) reduces to the Ricci curvature tensor itself,
\begin{equation}
 \label{electrovac_2}
   R_{\mu\nu} = \frac{\kappa}{2}\left( F_{\mu\rho} {F_{\nu}}^{\rho}
 + \widetilde{F}_{\mu\rho}{\widetilde{F}_{\nu}}\,^{\rho} \right).
\end{equation}
Since in the Minkowski frame $T_{00}=\frac{1}{2}(|{\bm E}|^2+ |{\bm B}|^2)$ is non-negative, we require that 
\begin{equation}
 \label{condicao_Rainich_2}
 R_{00}\geq 0,
\end{equation}
or $R_{\mu\nu}u^{\mu}u^{\nu} \geq 0$ for any time-like vector $u^{\mu} \Rightarrow u^{\mu}u_{\mu} = -1$.

We have shown that in order for symmetric tensor $T_{\mu\nu}$ to represent the energy-momentum tensor of an 
electromagnetic field, it must necessarily satisfy the relation (\ref{quadrado_de_T_2}), where we have
 ${T_{\mu\rho}}{T^{\rho\nu}} = \frac{1}{4} \left(I_1^2+ I_2^2\right)\,\,{\delta_{\mu}}^{\nu}$. 
 Now, we can obtain the same result to the Ricci curvature tensor. 
 Again, from equation (\ref{identidade_2}) we have that:
\begin{equation}
  F_{\mu\rho} {F_{\nu}}^{\rho} - \widetilde{F}_{\mu\rho}{\widetilde{F}_{\nu}}\,^{\rho} = 
  \frac{1}{2} g_{\mu\nu}F_{\rho\sigma}F^{\rho\sigma},\nonumber
\end{equation}
and using the above equation with (\ref{electrovac_2}) we have the following
equations system:
\begin{equation}
\begin{cases}
  R_{\mu\nu} = \dfrac{\kappa}{2}\left( F_{\mu\rho} {F_{\nu}}^{\rho}
 + \widetilde{F}_{\mu\rho}{\widetilde{F}_{\nu}}\,^{\rho} \right)\cr\cr
 \dfrac{\kappa}{4}\, g_{\mu\nu}F_{\rho\sigma}F^{\rho\sigma} = \dfrac{\kappa}{2}\, 
 \left( F_{\mu\rho} {F_{\nu}}^{\rho} - \widetilde{F}_{\mu\rho}{\widetilde{F}_{\nu}}\,^{\rho}\right).
\end{cases} \nonumber
\end{equation}
By adding together the above equations , we have:
\begin{equation}
 R_{\mu\nu}+  \frac{\kappa}{4}\, g_{\mu\nu}F_{\rho\sigma}F^{\rho\sigma} = \kappa F_{\mu\rho} {F_{\nu}}^{\rho},
 \nonumber
\end{equation}
and by subtraction we have:
\begin{equation}
 R_{\mu\nu} - \frac{\kappa}{4}\, g_{\mu\nu}F_{\rho\sigma}F^{\rho\sigma} = 
 \kappa \widetilde{F}_{\mu\rho}{\widetilde{F}_{\nu}}\,^{\rho}.\nonumber
\end{equation}
By contraction of two previous equations in order to get $R_{\mu\nu}R^{\nu\lambda}$, it follows that:
\begin{eqnarray}
 \left(R_{\mu\nu}+  \frac{\kappa}{4}\, g_{\mu\nu}F_{\rho\sigma}F^{\rho\sigma}\right)
  \left(R^{\nu\lambda} - \frac{\kappa}{4}\, g^{\nu\lambda}F_{\rho\sigma}F^{\rho\sigma}\right) &=&
  \kappa^2 F_{\mu\rho} {F_{\nu}}^{\rho}{\widetilde{F}^{\nu}}\,_{\sigma}{\widetilde{F}}^{\lambda\sigma}\cr
  R_{\mu\nu}R^{\nu\lambda} - \frac{k^2}{16}\,{\delta_{\mu}}^{\lambda}\left(F_{\rho\sigma}F^{\rho\sigma} \right)^2
  &=& \kappa^2 F_{\mu\rho} {\widetilde{F}}^{\lambda\sigma} ({F_{\nu}}^{\rho}{\widetilde{F}^{\nu}}\,_{\sigma}).\nonumber
\end{eqnarray}
From equation (\ref{identidade_5}) where we have that
\begin{equation}
\label{identidade_tensorial_Rainich_2}
{F_{\nu}}^{\rho}{\widetilde{F}^{\nu}}\,_{\sigma}=\frac{1}{4}\,\,{\delta^{\rho}}_{\sigma}
(\widetilde{F}_{\tau\phi}F^{\tau\phi}),
\end{equation}
we have
\begin{equation}
 R_{\mu\nu}R^{\nu\lambda} - \frac{k^2}{16}\,{\delta_{\mu}}^{\lambda}\left(F_{\rho\sigma}F^{\rho\sigma} \right)^2
  = \frac{\kappa^2}{4}\,  F_{\mu\rho} {\widetilde{F}}^{\lambda\sigma}\,
{\delta^{\rho}}_{\sigma}\,(\widetilde{F}_{\tau\phi}F^{\tau\phi})=  
\frac{\kappa^2}{4}\,  F_{\mu\rho} {\widetilde{F}}^{\lambda\rho}(\widetilde{F}_{\tau\phi}F^{\tau\phi}),\nonumber
\end{equation}
and again with (\ref{identidade_tensorial_Rainich_2}), 
\begin{equation}
 R_{\mu\nu}R^{\nu\lambda} - \frac{k^2}{16}\,{\delta_{\mu}}^{\lambda}\left(F_{\rho\sigma}F^{\rho\sigma} \right)^2
  =\frac{\kappa^2}{16} {\delta_{\mu}}^{\lambda} (\widetilde{F}_{\rho\sigma}F^{\rho\sigma})^2,\nonumber
\end{equation}
it results in
\begin{equation}
 \label{Rainich_3}
 R_{\mu\nu}R^{\nu\lambda} = \frac{\kappa^2}{16}\, {\delta_{\mu}}^{\lambda}\left[( F_{\rho\sigma}F^{\rho\sigma})^2
 + (\widetilde{F}_{\rho\sigma}F^{\rho\sigma})^2 \right].
\end{equation}
If we contract above equation over $\mu$ and $\lambda$, we find:
\begin{equation}
 \label{Rainich_4}
 R_{\mu\nu}R^{\mu\nu} = \frac{\kappa^2}{4} \left[( F_{\rho\sigma}F^{\rho\sigma})^2
 + (\widetilde{F}_{\rho\sigma}F^{\rho\sigma})^2 \right],
\end{equation}
Substituting above equation (\ref{Rainich_4}) into (\ref{Rainich_3}), we find the following condition:
\begin{equation}
  \label{Rainich_5}
 R_{\mu\lambda}R^{\lambda\nu} = \frac{1}{4}\, {\delta_{\mu}}^{\nu}\left(R_{\rho\sigma}R^{\rho\sigma}\right).
\end{equation}
Comparing the above result with (\ref{quadrado_de_T_3}), 
${T_{\mu\rho}}{T^{\rho\nu}} = \frac{1}{4} \,\,{\delta_{\mu}}^{\nu} \left(
 {T_{\kappa\lambda}}{T^{\kappa\lambda}}\right)$, we can observe that both expressions have the same algebraic
relationships.

Again, we may also recall from  from (\ref{invariante_1}) and (\ref{invariante_2}), where we have:
\begin{equation}
 I_1 = \frac{1}{2}F_{\rho\sigma}F^{\rho\sigma}= |{\bm B}|^2 - |{\bm E}|^2 \hspace*{1cm}
 \mbox{and} \hspace*{1cm} 
 I_2 =  \frac{1}{2}F_{\rho\sigma}\tilde{F}^{\rho\sigma} = -2{\bm E}\cdot{\bm B}.\nonumber
\end{equation}
Now, we obtain that (\ref{Rainich_4}) can be write as
\begin{equation}
 \label{Rainich_6}
 R_{\mu\nu}R^{\mu\nu} = {\kappa^2} \left[(|{\bm B}|^2 - |{\bm E}|^2)^2+ (2{\bm E}\cdot{\bm B})^2 \right]=
 {\kappa^2} \left[I_1^2  +I_2^2 \right],
\end{equation}
and the expression (\ref{Rainich_5}) is equivalent to
\begin{equation}
  \label{Rainich_7}
  R_{\mu\lambda}R^{\lambda\nu} = \frac{\kappa^2}{4}\, {\delta_{\mu}}^{\nu}\left(I_1^2  +I_2^2 \right).
\end{equation}

We have shown that in order for the Einstein-Maxwell electrovac equations the energy-momentum tensor
and the Ricci curvature tensor  must necessarily satisfy the following conditions:
\begin{eqnarray}
\label{Rainich_conditions}
 T={T_{\mu}}^{\mu} = 0 \hspace*{1cm} &\Longleftrightarrow& \hspace*{1cm} R={R_{\mu}}^{\mu} = 0 \cr
 T_{00}\geq 0 \hspace*{1cm} &\Longleftrightarrow& \hspace*{1cm} R_{00}\geq 0 \cr
{T_{\mu\rho}}{T^{\rho\nu}} = \frac{1}{4} \,\,{\delta_{\mu}}^{\nu} \left(
 {T_{\kappa\lambda}}{T^{\kappa\lambda}}\right)  \hspace*{1cm} &\Longleftrightarrow& \hspace*{1cm}
 R_{\mu\lambda}R^{\lambda\nu} = \frac{1}{4}\, {\delta_{\mu}}^{\nu}\left(R_{\rho\sigma}R^{\rho\sigma}\right) 
\end{eqnarray}
These relations are called: {\it Rainich algebraic conditions}. With these conditions,  we shall see 
that we can re-express the content of the Einstein-Maxwell electrovac
equations in a purely geometrical form. This program was carried out in 1925 by Rainich \cite{Rainich}
and in 1957 rediscovered by Misner and Wheeler \cite{Misner}. It has been called by Misner and Wheeler
the {\it already unified field theory} and this program was the basis of their {\it geometrodynamics}.

\subsubsection{The Ricci curvature tensor under duality rotations}

The Einstein-Maxwell electrovac equations from (\ref{electrovac_2}), where we have that
\begin{equation}
 R_{\mu\nu} =\frac{\kappa}{2}\left( F_{\mu\rho} {F_{\nu}}^{\rho}
 + \widetilde{F}_{\mu\rho}{\widetilde{F}_{\nu}}\,^{\rho} \right) =  
 \frac{\kappa}{2}\,{\cal F}_{\mu\rho}\left({{\cal F}_{\nu}}\,^{\rho}\right)^{*},\nonumber
\end{equation}
is invariant under duality rotation of electromagnetic field in accordance with (\ref{rotacao_dual_2.1}), where:
\begin{equation}
 {\cal F}_{\mu\nu} \rightarrow  {\cal F}_{\mu\nu} e^{-i\alpha}.\nonumber
\end{equation}
It is also instructive to use transformations (\ref{rotacao_dual_4}), where we have
\begin{eqnarray}
  F_{\mu\nu} &=& f_{\mu\nu}\,\cos{\alpha} - \widetilde{f}_{\mu\nu}\,\sin\alpha \cr
  \widetilde{F}_{\mu\nu} &=& f_{\mu\nu}\,\sin{\alpha} + \widetilde{f}_{\mu\nu}\,\cos\alpha, \nonumber
\end{eqnarray}
and verify  that:
\begin{equation}
 F_{\mu\nu}F^{\nu\rho}= f_{\mu\nu}f^{\nu\rho} \cos^2{\alpha} + \widetilde{f}_{\mu\nu}\widetilde{f}^{\nu\rho}
 \sin^2{\alpha} -\left(f_{\mu\nu}\widetilde{f}^{\nu\rho} +\widetilde{f}_{\mu\nu}f^{\nu\rho} \nonumber
 \right)\sin{\alpha}\cos{\alpha}
\end{equation}
and
\begin{equation}
 \widetilde{F}_{\mu\nu}\widetilde{F}^{\nu\rho} = 
 f_{\mu\nu}f^{\nu\rho} \sin^2{\alpha} + \widetilde{f}_{\mu\nu}\widetilde{f}^{\nu\rho}
 \cos^2{\alpha} + \left(f_{\mu\nu}\widetilde{f}^{\nu\rho} +\widetilde{f}_{\mu\nu}f^{\nu\rho} \nonumber
 \right)\sin{\alpha}\cos{\alpha},
\end{equation}
where the addition of two above equations results in:
\begin{equation}
  F_{\mu\nu}F^{\nu\rho} + \widetilde{F}_{\mu\nu}\widetilde{F}^{\nu\rho} = 
   f_{\mu\nu}f^{\nu\rho} + \widetilde{f}_{\mu\nu}\widetilde{f}^{\nu\rho}.\nonumber
\end{equation}
Thus, we have that the Einstein-Maxwell system of simultaneous equations, outside the electromagnetic sources,
is:
\begin{equation}
\label{electrovac_3}
 R_{\mu\nu}=\frac{\kappa}{2}\left( F_{\mu\rho} {F_{\nu}}^{\rho}
 + \widetilde{F}_{\mu\rho}{\widetilde{F}_{\nu}}\,^{\rho} \right) =\frac{\kappa}{2}\left( f_{\mu\rho} {f_{\nu}}^{\rho}
 + \widetilde{f}_{\mu\rho}{\widetilde{f}_{\nu}}\,^{\rho} \right),
\end{equation}
invariant under duality rotation of electromagnetic field.

\subsubsection{The Riemann curvature tensor of Einstein-Maxwell equations}
The Riemann curvature tensor, can be uniquely decomposed into parts:
\begin{equation}
\label{riemann_decomposicao}
 R_{\rho\sigma\mu\nu} = E_{\rho\sigma\mu\nu} +G_{\rho\sigma\mu\nu} +C_{\rho\sigma\mu\nu}
\end{equation}
where
\begin{equation}
 E_{\rho\sigma\mu\nu}=\frac{1}{2}(g_{\rho\mu}S_{\sigma\nu}+g_{\sigma\nu}S_{\rho\mu}
 -g_{\rho\nu}S_{\sigma\mu}-g_{\sigma\mu}S_{\rho\nu}),\nonumber
\end{equation}
with
\begin{equation}
 S_{\sigma\nu} = R_{\sigma\nu} -\frac{1}{4} g_{\sigma\nu} R, \nonumber
\end{equation}
\begin{equation}
 G_{\rho\sigma\mu\nu}=\frac{R}{12}\Big(g_{\rho\mu}g_{\sigma\nu}-g_{\rho\nu}g_{\sigma\mu}\Big),\nonumber
\end{equation}
$C_{\rho\sigma\mu\nu}$ is the Weyl conformal tensor with property ${C^{\mu}}_{\sigma\mu\nu}=C_{\sigma\nu}=0$.

Because Rainich condition, $R=0$ and because Weyl tensor is that part of curvature tensor such that all contractions
vanish, the relevant part of Riemann tensor (\ref{riemann_decomposicao}) is 
\begin{equation}
\label{tensor_E_1}
 E_{\rho\sigma\mu\nu}=\frac{1}{2}(g_{\rho\mu}R_{\sigma\nu}+g_{\sigma\nu}R_{\rho\mu}
 -g_{\rho\nu}R_{\sigma\mu}-g_{\sigma\mu}R_{\rho\nu}).
\end{equation}
The tensor $ E_{\rho\sigma\mu\nu}$ has two pair of bivector indices, so we can introduce the notions 
of right dual,
\begin{equation}
\label{right_dual}
   E^{\sim}_{\rho\sigma\mu\nu} =\frac{1}{2}\epsilon_{\mu\nu\tau\phi}{E_{\rho\sigma}}^{\tau\phi}
\end{equation}
and left dual,
\begin{equation}
\label{left_dual}
  ^{\sim} E_{\rho\sigma\mu\nu} =\frac{1}{2}\epsilon_{\rho\sigma\tau\phi}{E^{\tau\phi}}_{\mu\nu}.
\end{equation}
It turns out that left dual tensor and right dual tensor obey the relation:
\begin{equation}
\label{identidade_dual_1}
  ^{\sim} E_{\rho\sigma\mu\nu} = - E^{\sim}_{\rho\sigma\mu\nu}.
\end{equation}

We can state that the tensor  $E_{\rho\sigma\mu\nu}$ can be expressed as a combination of
bivectors $f_{\mu\nu}$ and $\widetilde{f}_{\rho\sigma}$, such as:
\begin{equation}
 \label{curvatura_eletromagnética_9}
 E_{\mu\nu\rho\sigma} = \frac{\kappa}{2} \left( f_{\mu\nu}f_{\rho\sigma}+f_{\mu\nu}\widetilde{f}_{\rho\sigma}+
 \widetilde{f}_{\mu\nu}f_{\rho\sigma}+\widetilde{f}_{\mu\nu}\widetilde{f}_{\rho\sigma}\right).
\end{equation}
Accordingly with (\ref{right_dual}), the right dual operation in above expression $ E_{\mu\nu\rho\sigma}$, we 
have
\begin{eqnarray}
E^{\sim}_{\rho\sigma\mu\nu}  &=&\frac{\kappa}{2}\,\,\frac{1}{2}\epsilon_{\mu\nu\tau\phi} (f_{\rho\sigma}f^{\tau\phi} +
f_{\rho\sigma}\widetilde{f}^{\tau\phi} + \widetilde{f}_{\rho\sigma}f^{\tau\phi} + 
\widetilde{f}_{\rho\sigma}\widetilde{f}^{\tau\phi}),\nonumber
\end{eqnarray}
we have $\frac{1}{2}\epsilon_{\mu\nu\tau\phi}f^{\tau\phi} = \widetilde{f}_{\mu\nu}$ and 
$\frac{1}{2}\epsilon_{\mu\nu\tau\phi}\widetilde{f}^{\tau\phi} = -{f}_{\mu\nu}$ thus,
the above expression results in:
\begin{equation}
\label{right_dual_2}
E^{\sim}_{\rho\sigma\mu\nu} = \frac{\kappa}{2}(f_{\rho\sigma}\widetilde{f}_{\mu\nu} -
f_{\rho\sigma}f_{\mu\nu} + \widetilde{f}_{\rho\sigma}\widetilde{f}_{\mu\nu} - 
\widetilde{f}_{\rho\sigma}f_{\mu\nu}).
\end{equation}
The left dual operation results in
\begin{equation}
\label{left_dual_2}
 ^{\sim} E_{\rho\sigma\mu\nu} = \frac{\kappa}{2} (\widetilde{f}_{\rho\sigma}f_{\mu\nu} 
+ \widetilde{f}_{\rho\sigma}\widetilde{f}_{\mu\nu} - f_{\rho\sigma}f_{\mu\nu} -
f_{\rho\sigma}\widetilde{f}_{\mu\nu}).
\end{equation}
In accordance with relation (\ref{identidade_dual_1}) we have
\begin{eqnarray}
 f_{\rho\sigma}\widetilde{f}_{\mu\nu} -
f_{\rho\sigma}f_{\mu\nu} + \widetilde{f}_{\rho\sigma}\widetilde{f}_{\mu\nu} - 
\widetilde{f}_{\rho\sigma}f_{\mu\nu} &=& - (\widetilde{f}_{\rho\sigma}f_{\mu\nu} 
+ \widetilde{f}_{\rho\sigma}\widetilde{f}_{\mu\nu} - f_{\rho\sigma}f_{\mu\nu} -
f_{\rho\sigma}\widetilde{f}_{\mu\nu})\cr
(f_{\rho\sigma}\widetilde{f}_{\mu\nu} - \widetilde{f}_{\rho\sigma}f_{\mu\nu}) +
(\widetilde{f}_{\rho\sigma}\widetilde{f}_{\mu\nu}-f_{\rho\sigma}f_{\mu\nu}) &=&
(f_{\rho\sigma}\widetilde{f}_{\mu\nu} - \widetilde{f}_{\rho\sigma}f_{\mu\nu}) -
(\widetilde{f}_{\rho\sigma}\widetilde{f}_{\mu\nu}-f_{\rho\sigma}f_{\mu\nu}),\nonumber
\end{eqnarray}
that results 
\begin{equation}
 \widetilde{f}_{\rho\sigma}\widetilde{f}_{\mu\nu}-f_{\rho\sigma}f_{\mu\nu} = 0.\nonumber
\end{equation}
Consequently we have from (\ref{right_dual_2}) and  (\ref{left_dual_2}) that:
\begin{equation}
E^{\sim}_{\rho\sigma\mu\nu} = \frac{\kappa}{2}(f_{\rho\sigma}\widetilde{f}_{\mu\nu} - 
\widetilde{f}_{\rho\sigma}f_{\mu\nu}) \hspace*{1cm} \mbox{and} \hspace*{1cm}
 ^{\sim} E_{\rho\sigma\mu\nu} = \frac{\kappa}{2} (\widetilde{f}_{\rho\sigma}f_{\mu\nu} 
-f_{\rho\sigma}\widetilde{f}_{\mu\nu}). \nonumber
\end{equation}
We note that one more left dual operation on $^{\sim} E_{\rho\sigma\mu\nu}$ results in:
\begin{equation}
  ^{\sim\sim} E_{\rho\sigma\mu\nu} = -\frac{\kappa}{2} (\widetilde{f}_{\rho\sigma}\widetilde{f}_{\mu\nu} 
-\widetilde{\widetilde{f}}_{\rho\sigma}f_{\mu\nu})= 
-\frac{\kappa}{2} (\widetilde{f}_{\rho\sigma}\widetilde{f}_{\mu\nu} 
+f_{\rho\sigma}f_{\mu\nu}) = - E_{\rho\sigma\mu\nu}, \nonumber
\end{equation}
where we recall (\ref{dual_dual_bivector}),\, ${\widetilde{\!\widetilde{f}}}_{\rho\sigma}=-f_{\rho\sigma}$. Then
 the tensor $ E_{\rho\sigma\mu\nu}$ has the form:
\begin{equation}
 \label{curvatura_eletromagnética_10}
 E_{\rho\sigma\mu\nu} = \frac{\kappa}{2} \left(\widetilde{f}_{\rho\sigma}\widetilde{f}_{\mu\nu} 
+f_{\rho\sigma}f_{\mu\nu}\right). 
\end{equation}

Having determined $E_{\rho\sigma\mu\nu}$, we can find Einstein-Maxwell electrovac equation by contraction of 
Riemann tensor (\ref{riemann_decomposicao}), where we have
\begin{equation}
 R_{\mu\nu} = E_{\mu\nu} = \frac{\kappa}{2}\left(f_{\mu\rho}{f_{\nu}}^{\rho} + 
 \widetilde{f}_{\mu\rho}{\widetilde{f}_{\nu}}\,^{\rho}\right)
\end{equation}

In order to find the complexion, the angle $\alpha$ of duality rotation (\ref{rotacao_dual_3}), 
we need some calculations.
We can evaluate covariant derivative into two equations from (\ref{rotacao_dual_3}), remembering the Maxwell's
equations without sources, $\nabla_{\nu}F^{\mu\nu}=0$,
\begin{eqnarray}
 \nabla_{\nu} f^{\mu\nu} &=& \nabla_{\nu}(F^{\mu\nu}\,\cos{\alpha} + \widetilde{F}^{\mu\nu}\,\sin\alpha)
 = (-F^{\mu\nu}\,\sin{\alpha} + \widetilde{F}^{\mu\nu}\,\cos{\alpha})\partial_{\nu}\alpha
 = \widetilde{f}^{\mu\nu} \partial_{\nu}\alpha \cr
\nabla_{\nu}  \widetilde{f}^{\mu\nu} &=& \nabla_{\nu}(-F^{\mu\nu}\,\sin{\alpha} + \widetilde{F}^{\mu\nu}\,\cos\alpha)
= (F^{\mu\nu}\,\cos{\alpha} + \widetilde{F}^{\mu\nu}\,\sin\alpha)\partial_{\nu}\alpha 
= - f^{\mu\nu} \partial_{\nu}\alpha, \nonumber
\end{eqnarray}
we can contract,
\begin{eqnarray}
 \widetilde{f}_{\rho\mu} \nabla_{\nu} f^{\mu\nu} &=&   
 \widetilde{f}_{\rho\mu}\widetilde{f}^{\mu\nu} \partial_{\nu}\alpha \cr
 f_{\rho\mu} \nabla_{\nu}  \widetilde{f}^{\mu\nu} &=& - f_{\rho\mu}  f^{\mu\nu} \partial_{\nu}\alpha, \nonumber
\end{eqnarray}
where the sum of two above equations results in:
\begin{equation}
 \widetilde{f}_{\rho\mu} \nabla_{\nu} f^{\mu\nu} + f_{\rho\mu} \nabla_{\nu}  \widetilde{f}^{\mu\nu} =
 - ( f_{\rho\mu}  f^{\mu\nu} - \widetilde{f}_{\rho\mu}\widetilde{f}^{\mu\nu}) \partial_{\nu}\alpha,\nonumber
\end{equation}
and with equation (\ref{identidade_2}) we have:
$f_{\rho\mu}  f^{\nu\mu} - \widetilde{f}_{\rho\mu}\widetilde{f}^{\nu\mu}= 
\frac{1}{2}{\delta_{\rho}}^{\nu}  f_{\kappa\lambda}f^{\kappa\lambda} $,
such as :
\begin{equation}
 \widetilde{f}_{\rho\mu} \nabla_{\nu} f^{\mu\nu} + f_{\rho\mu} \nabla_{\nu}  \widetilde{f}^{\mu\nu} =
 \frac{1}{2}{\delta_{\rho}}^{\nu}  f_{\kappa\lambda}f^{\kappa\lambda} \partial_{\nu}\alpha, \nonumber
\end{equation}
and we can define the gradient $\partial_{\nu}\alpha = \alpha_{\nu}$,
\begin{equation}
 \label{complexion_1}
 \alpha_{\rho} = 2  \left(f_{\kappa\lambda}f^{\kappa\lambda}\right)^{-1}
 \left(\widetilde{f}_{\rho\mu} \nabla_{\nu} f^{\mu\nu} + f_{\rho\mu} \nabla_{\nu}  \widetilde{f}^{\mu\nu}\right).
\end{equation}
In order to express the gradient $\partial_{\nu}\alpha$ in terms of geometrical quantities, we can use 
(\ref{curvatura_eletromagnética_10}),
\begin{equation}
 \nabla_{\nu}E^{\rho\sigma\mu\nu} = \frac{\kappa}{2}\left[ \Big(\nabla_{\nu}f^{\rho\sigma}\Big)f^{\mu\nu}
 + f^{\rho\sigma} \nabla_{\nu}f^{\mu\nu}+ \left(\nabla_{\nu}\widetilde{f}^{\rho\sigma}\right)
 \widetilde{f}^{\mu\nu}+ \widetilde{f}^{\rho\sigma} \nabla_{\nu}  \widetilde{f}^{\mu\nu}\right],\nonumber
\end{equation}
and
\begin{equation}
 E^{\sim}_{\tau\mu\rho\sigma} = \frac{\kappa}{2}\left(f_{\tau\mu}\widetilde{f}_{\rho\sigma} - 
 \widetilde{f}_{\tau\mu}f_{\rho\sigma}\right). \nonumber
\end{equation}
Then we can contract two above equations,
\begin{eqnarray}
  \big(\nabla_{\nu}E^{\rho\sigma\mu\nu}\big) E^{\sim}_{\tau\mu\rho\sigma} &=& \frac{\kappa^2}{4}\Big[
\Big(\nabla_{\nu}f^{\rho\sigma}\Big)f^{\mu\nu} f_{\tau\mu}\widetilde{f}_{\rho\sigma} -
\Big(\nabla_{\nu}f^{\rho\sigma}\Big)f^{\mu\nu} \widetilde{f}_{\tau\mu}f_{\rho\sigma} +
 f^{\rho\sigma} \Big(\nabla_{\nu}f^{\mu\nu}\Big)f_{\tau\mu}\widetilde{f}_{\rho\sigma}
 -  f^{\rho\sigma}\Big( \nabla_{\nu}f^{\mu\nu}\Big)\widetilde{f}_{\tau\mu}f_{\rho\sigma}\cr 
 & & +\left(\nabla_{\nu}\widetilde{f}^{\rho\sigma}\right)\widetilde{f}^{\mu\nu}f_{\tau\mu}\widetilde{f}_{\rho\sigma}
 - \left(\nabla_{\nu}\widetilde{f}^{\rho\sigma}\right)\widetilde{f}^{\mu\nu}\widetilde{f}_{\tau\mu}f_{\rho\sigma}
 + \widetilde{f}^{\rho\sigma} \Big(\nabla_{\nu}  \widetilde{f}^{\mu\nu}\Big) f_{\tau\mu}\widetilde{f}_{\rho\sigma}
 - \widetilde{f}^{\rho\sigma} \Big(\nabla_{\nu}  \widetilde{f}^{\mu\nu}\Big)\widetilde{f}_{\tau\mu}f_{\rho\sigma}
\Big].\nonumber
\end{eqnarray}
Because extremal field condition, $f_{\mu\nu}\widetilde{f}^{\mu\nu} =0$, we have from (\ref{identidade_5}) that
\begin{equation}
 {f_{\nu}}^{\rho}{\widetilde{f}^{\nu}}\,_{\sigma} =\frac{1}{4}{\delta^{\rho}}_{\sigma}\Big(\widetilde{f}_{\mu\nu}
 f^{\mu\nu}\Big) = 0, \nonumber
\end{equation}
from which we find that
\begin{equation}
   E^{\sim}_{\tau\mu\rho\sigma} \nabla_{\nu}E^{\rho\sigma\mu\nu} = \frac{\kappa^2}{4}\Big[
\widetilde{f}_{\rho\sigma} \Big(\nabla_{\nu}f^{\rho\sigma}\Big)f^{\mu\nu} f_{\tau\mu}-
 f^{\rho\sigma}f_{\rho\sigma} \widetilde{f}_{\tau\mu} \nabla_{\nu}f^{\mu\nu}
 -f_{\rho\sigma} \left(\nabla_{\nu}\widetilde{f}^{\rho\sigma}\right)\widetilde{f}^{\mu\nu}\widetilde{f}_{\tau\mu}
 + \widetilde{f}^{\rho\sigma}\widetilde{f}_{\rho\sigma}f_{\tau\mu} \nabla_{\nu}  \widetilde{f}^{\mu\nu} 
\Big],\nonumber
\end{equation}
with $\widetilde{f}^{\rho\sigma}\widetilde{f}_{\rho\sigma} = - f^{\rho\sigma}f_{\rho\sigma} $ we have:
\begin{equation}
\label{complexion_2}
   E^{\sim}_{\tau\mu\rho\sigma} \nabla_{\nu}E^{\rho\sigma\mu\nu} = \frac{\kappa^2}{4}\Big[
-  \Big(f^{\rho\sigma}f_{\rho\sigma}\Big)\Big(\widetilde{f}_{\tau\mu}  \nabla_{\nu}f^{\mu\nu} 
+f_{\tau\mu} \nabla_{\nu}  \widetilde{f}^{\mu\nu}\Big)
+\widetilde{f}_{\rho\sigma} \Big(\nabla_{\nu}f^{\rho\sigma}\Big)f^{\mu\nu} f_{\tau\mu}
 -f_{\rho\sigma} \left(\nabla_{\nu}\widetilde{f}^{\rho\sigma}\right)\widetilde{f}^{\mu\nu}\widetilde{f}_{\tau\mu}
\Big].
\end{equation}
Observe that:
\begin{equation}
\label{identidade_6}
 \widetilde{f}_{\rho\sigma}\nabla_{\nu} f^{\rho\sigma}=
 \frac{1}{2}\epsilon_{\rho\sigma\kappa\lambda}f^{\kappa\lambda}\nabla_{\nu}f^{\rho\sigma} 
 = f^{\kappa\lambda}\nabla_{\nu}\left(\frac{1}{2}\epsilon_{\rho\sigma\kappa\lambda} f^{\rho\sigma}\right)
 =  f^{\kappa\lambda}\nabla_{\nu}\left(\frac{1}{2}\epsilon_{\kappa\lambda\rho\sigma} f^{\rho\sigma}\right)
 = f^{\kappa\lambda}\nabla_{\nu}\widetilde{f}_{\kappa\lambda}.
\end{equation}
And from extremal field condition $f_{\mu\nu}\widetilde{f}^{\mu\nu} =0$ we have
\begin{equation}
 \nabla_{\nu}\left(\widetilde{f}_{\rho\sigma}f^{\rho\sigma}\right) = 
 \left(\nabla_{\nu}\widetilde{f}_{\rho\sigma}\right)f^{\rho\sigma} + 
 \widetilde{f}_{\rho\sigma}\nabla_{\nu}f^{\rho\sigma} = 0, \nonumber
\end{equation}
or
\begin{equation}
\label{identidade_7}
 \widetilde{f}_{\rho\sigma}\nabla_{\nu}f^{\rho\sigma} = -f^{\rho\sigma}\nabla_{\nu}\widetilde{f}_{\rho\sigma}.
\end{equation}
Adding (\ref{identidade_6}) and (\ref{identidade_7}),
we have that:
\begin{equation}
  \widetilde{f}_{\rho\sigma}\nabla_{\nu} f^{\rho\sigma}=0.\nonumber
\end{equation}
So, with this, the equation (\ref{complexion_2}) reduces to
\begin{equation}
   E^{\sim}_{\tau\mu\rho\sigma} \nabla_{\nu}E^{\rho\sigma\mu\nu} = -\frac{\kappa^2}{4}
 \Big(f^{\rho\sigma}f_{\rho\sigma}\Big)\Big(\widetilde{f}_{\tau\mu}  \nabla_{\nu}f^{\mu\nu} 
+f_{\tau\mu} \nabla_{\nu}  \widetilde{f}^{\mu\nu}\Big), \nonumber
\end{equation}
and dividing by $\Big(f^{\rho\sigma}f_{\rho\sigma}\Big)^2$ we find that
\begin{equation}
 \label{complexion_3}
 \frac{ \widetilde{f}_{\tau\mu}  \nabla_{\nu}f^{\mu\nu} +f_{\tau\mu} \nabla_{\nu}  \widetilde{f}^{\mu\nu}}
 {\Big(f^{\rho\sigma}f_{\rho\sigma}\Big)} = -\frac{4\, E^{\sim}_{\tau\mu\rho\sigma} \nabla_{\nu}E^{\rho\sigma\mu\nu}}
 {\kappa^2\Big(f^{\rho\sigma}f_{\rho\sigma}\Big)^2}.
\end{equation}
From expression for gradient of complexion $\alpha$ (\ref{complexion_1}) we have:
\begin{equation}
 \label{complexion_4}
 \alpha_{\tau} = - 2 \left(\frac{2}{\kappa f^{\rho\sigma}f_{\rho\sigma}}\right)^2
  E^{\sim}_{\tau\mu\rho\sigma} \nabla_{\nu}E^{\rho\sigma\mu\nu}.
\end{equation}

The next step is to rewrite the complexion vector $\alpha_{\tau}$  in terms of the Ricci curvature tensor.
The contracted Ricci curvature tensor given by (\ref{Rainich_4}) can be given
in terms of $f_{\mu\nu}$, if we use (\ref{condicao_extrema_3}):
\begin{equation}
\label{electrovac_4}
 R_{\mu\nu}R^{\mu\nu} = \frac{\kappa^2}{4} \left( f_{\mu\nu}f^{\mu\nu}\right)^2,
\end{equation}
in this case the complexion vector $\alpha_{\tau}$ is
\begin{equation}
 \label{complexion_5}
 \alpha_{\tau} = - 2 \left(R_{\rho\sigma}R^{\rho\sigma}\right)^{-1}
  E^{\sim}_{\tau\mu\rho\sigma} \nabla_{\nu}E^{\rho\sigma\mu\nu}.
\end{equation}
The covariant derivative applied on tensor (\ref{tensor_E_1}) is:
\begin{equation}
 \nabla_{\nu}E^{\rho\sigma\mu\nu} = \frac{1}{2} \left( g^{\rho\mu}\nabla_{\nu}R^{\sigma\nu}
 +g^{\sigma\nu}\nabla_{\nu}R^{\rho\mu} - g^{\rho\nu}\nabla_{\nu}R^{\sigma\mu} 
 - g^{\sigma\mu}\nabla_{\nu}R^{\rho\nu}\right), \nonumber
\end{equation}
where the Einstein field equation satisfies
$\nabla_{\nu}G^{\mu\nu} = 0$, then with (\ref{condicao_Rainich_1}) we have $\nabla_{\nu}R^{\mu\nu} = 0$
and consequently we have that:
\begin{equation}
 \nabla_{\nu}E^{\rho\sigma\mu\nu} = \frac{1}{2} \left( \nabla^{\sigma}R^{\rho\mu} - \nabla^{\rho}R^{\sigma\mu} 
 \right).\nonumber
\end{equation}
Now from right dual operation on tensor $E_{\rho\sigma\mu\nu}$ (\ref{right_dual}), we have:
\begin{equation}
   E^{\sim}_{\tau\mu\rho\sigma} =\frac{1}{2}\epsilon_{\rho\sigma\kappa\lambda}{E_{\tau\mu}}^{\kappa\lambda}
   =\frac{1}{4}\epsilon_{\rho\sigma\kappa\lambda}\left({\delta_{\tau}}^{\kappa}{R_{\mu}}^{\lambda}+
  {\delta_{\mu}}^{\lambda}{R_{\tau}}^{\kappa} - {\delta_{\tau}}^{\lambda}{R_{\mu}}^{\kappa} 
  -  {\delta_{\mu}}^{\kappa}{R_{\tau}}^{\lambda}\right) = 
  \frac{1}{2}\left(\epsilon_{\rho\sigma\tau\lambda}{R_{\mu}}^{\lambda}  - 
  \epsilon_{\rho\sigma\mu\lambda}{R_{\tau}}^{\lambda}\right).\nonumber
\end{equation}
Thus we find that
\begin{equation}
 E^{\sim}_{\tau\mu\rho\sigma} \nabla_{\nu}E^{\rho\sigma\mu\nu} = 
 \frac{1}{2} \left( \nabla^{\sigma}R^{\rho\mu} - \nabla^{\rho}R^{\sigma\mu}  \right) \cdot
  \frac{1}{2}\left(\epsilon_{\rho\sigma\tau\lambda}{R_{\mu}}^{\lambda}  - 
  \epsilon_{\rho\sigma\mu\lambda}{R_{\tau}}^{\lambda}\right) =
  \frac{1}{4} \epsilon_{\rho\sigma\tau\lambda}{R_{\mu}}^{\lambda} 
  \left( \nabla^{\sigma}R^{\rho\mu} - \nabla^{\rho}R^{\sigma\mu}  \right)\nonumber
\end{equation}
or
\begin{equation}
 E^{\sim}_{\tau\mu\rho\sigma} \nabla_{\nu}E^{\rho\sigma\mu\nu} = 
 \frac{1}{2} \epsilon_{\rho\sigma\tau\lambda}{R_{\mu}}^{\lambda} \nabla^{\sigma} R^{\rho\mu}.\nonumber
\end{equation}
Putting this result into (\ref{complexion_5}) yields the desired result
\begin{equation}
 \label{complexion_6}
 \alpha_{\tau} = \left(R_{\mu\nu}R^{\mu\nu}\right)^{-1}
   \epsilon_{\tau\lambda\rho\sigma}{R_{\kappa}}^{\lambda}\nabla^{\rho}R^{\kappa\sigma}.
\end{equation}
In the words of C.W. Misner and J.A. Wheeler \cite{Misner}: ``The vector $\alpha_\tau$ of
(\ref{complexion_6}) has a well defined existence in any Riemannian space where the Ricci curvature tensor
$R_{\mu\nu}$ is non-null and differentiable. From such general Riemannian spaces the geometry of the 
Einstein-Maxwell theory is distinguished by the circumstance that this vector is not arbitrary,
but is the gradient of the complexion, $\alpha$, of the electromagnetic field.'' Consequently it follows that
the curl of $\alpha_{\tau}$ must vanish:
\begin{equation}
 \label{curl_complexion}
\nabla_{\sigma}\alpha_{\tau} -  \nabla_{\tau}\alpha_{\sigma} =0. 
\end{equation}
Louis Witten \cite{Witten} called the above equation the Rainich, Misner, Wheeler (RMW) differential condition.
If the RMW differential condition is satisfied, we can find the complexion $\alpha$ from (\ref{rotacao_dual_1})
by an integration
\begin{equation}
 \label{complexion_7}
 \alpha(x) = \int \alpha_{\tau} dx^{\tau} +\alpha_{0}.
\end{equation}
Thus the complexion $\alpha(x)$ is completely determined by (\ref{complexion_7}) up to an additive constant
provided that in the region of space under consideration Maxwell's equations hold everywhere (no singularities),
provided that the region of space is simply connected, and provided the line of integration not include any point 
where $R_{\mu\nu}R^{\mu\nu}=0$. For a multiply connected space or one in which there are singularities 
the appropriate demand on $\alpha_{\tau}$  is that the line integral of  $\alpha_{\tau}$  around any closed path 
shall be an integral multiple of $2\pi$,
\begin{equation}
 \oint \alpha_{\tau} dx^{\tau} = n\, 2\pi, \nonumber
\end{equation}
provided the line of integration does not touch any null points. This condition, plus the algebraic requirements 
of Rainich:
\begin{equation}
 {R_{\mu}}^{\mu} =0; \hspace*{1cm} R_{00}> 0; \hspace*{1cm} {R_{\rho}}^{\tau} {R_{\tau}}^{\sigma} = 
 \frac{1}{4}{\delta_{\rho}}^{\sigma} R_{\mu\nu}R^{\mu\nu} \nonumber
\end{equation}
gives the necessary and sufficient condition that Riemannian geometry shall reproduce the physics of Einstein
and Maxwell, provided that the Ricci curvature tensor $R_{\mu\nu}$ is non-null.

\section{Null Electromagnetic Field}

If we have
\begin{equation}
 R_{\mu\lambda}R^{\lambda\nu} = 0 \hspace*{1cm} \mbox{or} \hspace*{1cm}  
 T_{\mu\lambda}T^{\lambda\nu} = 0, \nonumber
\end{equation}
but the Ricci tensor which is not identically zero, then in this case 
we have null electromagnetic field, in accordance with (\ref{Rainich_6}) where the invariants 
$I_1$ and $I_2$ are vanished, such as
\begin{equation}
|{\bm B}|^2 = |{\bm E}|^2  \hspace*{1cm} \mbox {e}  \hspace*{1cm} 
{\bm E}\cdot{\bm B} = 0 \Rightarrow {\bm E}\perp {\bm B}.\nonumber
\end{equation}
Here the value for the angle $\alpha$, the complexion of electromagnetic field, becomes indeterminate, since
the angle $\alpha$ is evaluated from equation (\ref{condicao_extrema_1}).
The null electromagnetic field occurs most frequently in plane electromagnetic wave in vacuum.

Usually in curved spacetime, the tensors are expressed in a coordinate basis. However, for many purposes it is more
convenient to use an orthonormal basis or Minkowski reference system, where the components of the metric
tensor have the relationships with Minkowski metric given by
\begin{equation}
 {e_{\alpha}}^{\mu}{e_{\beta}}^{\nu}\, g_{\mu\nu} = \eta_{\alpha\beta}.\nonumber
\end{equation}
The coefficients ${e_{\alpha}}^{\mu}$ are the vierbeins that rotates the coordinate basis to an orthonormal basis.
Thus, the Ricci curvature tensor in Minkowski reference system is given by:
\begin{equation}
 R_{\alpha\beta} =  {e_{\alpha}}^{\mu}{e_{\alpha}}^{\nu} R_{\mu\nu}.\nonumber
\end{equation}
The Einstein-Maxwell electrovac equation (\ref{electrovac_2}), where 
$R_{\alpha\beta} = \kappa\, T_{\alpha\beta}$ in orthonormal basis,
help us to obtain the Ricci tensor from
the energy-momentum tensor  $T_{\alpha\beta}$ given by 
(\ref{tensor_energia_momento_eletromagnetico_3}),
\begin{equation}
 (R_{\alpha\beta})= \kappa(T_{\alpha\beta})= \kappa
 \begin{pmatrix}
  \frac{1}{2}(|{\bm E}|^2 +|{\bm B}|^2) & -({\bm E}\times{\bm B})\cr
  -({\bm E}\times{\bm B}) & \sigma_{ij}
 \end{pmatrix},\nonumber
\end{equation}
and make such a rotation in 3-space as will diagonalize the the $3\times 3$, space-space part  $\sigma_{ij}$, 
\begin{equation}
 (\sigma_{ij}) =
 \begin{pmatrix}
  \sigma_{xx} & 0 & 0 \cr
  0 & \sigma_{yy} & 0 \cr
  0 & 0 & \sigma_{zz}
 \end{pmatrix}. \nonumber
\end{equation}
When the electromagnetic field is null, $R_{\mu\lambda}R^{\lambda\nu} = 0 $, 
in Minkowski reference system we have
\begin{equation}
\eta^{\gamma\delta} R_{\alpha\gamma}R_{\beta\delta} = 0,\nonumber 
\end{equation}
then with $R_{12}=R_{13}=R_{23} =0 $, we have that:
\begin{enumerate}
 \item for $\alpha=\beta=0$ results in:
 \begin{equation}
  -(R_{00})^2 + (R_{01})^2 + (R_{02})^2 + (R_{03})^2 = 0; \nonumber
 \end{equation}
 \item  for $\alpha=0$ e $\beta=1$,
 \begin{equation}
  -R_{00}R_{01} + R_{01}R_{11} = 0; \nonumber
 \end{equation}
  \item  for $\alpha=0$ e $\beta=2$,
 \begin{equation}
  -R_{00}R_{02} + R_{02}R_{22} = 0; \nonumber
 \end{equation}
 \item  for $\alpha=0$ e $\beta=3$,
 \begin{equation}
  -R_{00}R_{03} + R_{03}R_{33} = 0; \nonumber
 \end{equation}
\item  for $\alpha=1$ e $\beta=2$, 
 \begin{equation}
  -R_{10}R_{02} = 0; \nonumber
 \end{equation}
\item  for $\alpha=1$ e $\beta=3$, 
 \begin{equation}
  -R_{10}R_{03} = 0; \nonumber
 \end{equation} 
\item  for $\alpha=2$ e $\beta=3$, 
 \begin{equation}
  -R_{20}R_{03} = 0; \nonumber
 \end{equation} 
\noindent For to the items 5, 6 e 7 we can choose $R_{02}=R_{03}=0$ with only $R_{01}\neq 0$. This choice,
$R_{01}\neq 0$ results in $R_{00}=R_{11}$ to the equation from item 2.
 \item  for $\alpha=1$ e $\beta=1$, 
 \begin{equation}
  -(R_{10})^2+(R_{11})^2 = 0 \hspace*{0.5cm}\Rightarrow  \hspace*{0.5cm} R_{10} = \pm\, R_{11} ; \nonumber
 \end{equation} 
 \item  for $\alpha=2$ e $\beta=2$, 
 \begin{equation}
  -(R_{20})^2+(R_{22})^2 = 0 , \nonumber
 \end{equation} 
 but with above choice  $R_{20}=0$, we have $R_{22}=0$.
 \item  For $\alpha=3$ e $\beta=3$, 
 \begin{equation}
  -(R_{30})^2+(R_{33})^2 = 0 , \nonumber
 \end{equation} 
in the same way, with the choice $R_{30}=0$, we have $R_{33}=0$.
 \end{enumerate}
We can recall equation from item 1, where we have $R_{00}=\pm R_{01}$.
The choice of sign can be the minus sign $R_{00}=- R_{01}$.
The Rainich conditions (\ref{Rainich_conditions}) require that  $R_{00} = \kappa T_{00} \geq 0$, where 
the energy density must be positive.
Putting $R_{00}={\kappa}{\epsilon^2}$, where $\kappa=\dfrac{8\pi G}{c^4}$,
the Ricci curvature tensor must have the form
\begin{equation}
 \label{tensor_Ricci_Campo_nulo}
 (R_{\alpha\beta}) ={\kappa}
 \begin{pmatrix}
  \epsilon^2 & -\epsilon^2 & 0 & 0 \cr
  -\epsilon^2 & \epsilon^2 & 0 & 0 \cr
  0 & 0 & 0 & 0 \cr
  0 & 0 & 0 & 0 \cr
 \end{pmatrix}
 ={\kappa\,\epsilon^2}
 \begin{pmatrix}
  1 & -1 & 0 & 0 \cr
  -1 & 1 & 0 & 0 \cr
  0 & 0 & 0 & 0 \cr
  0 & 0 & 0 & 0 \cr
 \end{pmatrix} 
\end{equation}
We can define a null vector:
\begin{equation}
 k_{\alpha} = (-1,1, 0, 0),
\end{equation}
where $ k_{\alpha}k^{\alpha} = k^2 = 0$. Therefore, the null electromagnetic field
has the Ricci curvature tensor in the form:
\begin{equation}
\label{Ricci_nulo}
 R_{\alpha\beta} = {\kappa\,\epsilon^2}\, k_{\alpha}k_{\beta}.
\end{equation}
The energy-momentum tensor of null electromagnetic field is:
\begin{equation}
\label{null_dust_tensor}
 T_{\alpha\beta} = \epsilon^2\,  k_{\alpha}k_{\beta}.
\end{equation}
Being covariant and true in one reference system, the decomposition (\ref{Ricci_nulo})
is valid in any reference system.

There is no Lorentz transformation that will diagonalize a null Ricci tensor (\ref{Ricci_nulo}), and
consequently the energy-momentum tensor $T_{\alpha\beta}$. Moreover,
there is no Lorentz transformation that will make a null vector timelike, or parallelize
field vectors ${\bm E}$ and ${\bm B}$ that satisfy the null condition, 
\begin{equation}
|{\bm B}|^2 = |{\bm E}|^2  \hspace*{1cm} \mbox {e}  \hspace*{1cm} 
{\bm E}\cdot{\bm B} = 0 \Rightarrow {\bm E}\perp {\bm B}.\nonumber
\end{equation}
The electromagnetic field-strength that satisfies the null condition is:
\begin{equation}
\label{field_strenght_1}
  (F_{\alpha\beta})=
 \begin{pmatrix}
  0 & 0 & -E_y & 0 \cr
   0 & 0 & B_z & 0 \cr
   E_y & -B_z & 0 & 0 \cr
   0 & 0 & 0 & 0 \cr
 \end{pmatrix} \hspace*{1cm} \mbox{with} \hspace*{1cm}
  (\tilde{F}_{\alpha\beta})=
 \begin{pmatrix}
  0 & 0 & 0 & -B_z \cr
  0 & 0 & 0 & E_y \cr
  0 & 0 & 0 & 0 \cr
  B_z & -E_y & 0 & 0 \cr
 \end{pmatrix}.
\end{equation}
The energy-momentum tensor of electromagnetic field from (\ref{tensor_energia_momento_eletromagnetico_2}) is
\begin{equation}
    T_{\mu\nu} = \frac{1}{2}\left( F_{\mu\rho} {F_{\nu}}^{\rho}
 + \widetilde{F}_{\mu\rho}{\widetilde{F}_{\nu}}\,^{\rho} \right),\nonumber
\end{equation}
that results to Ricci tensor:
\begin{equation}
 (T_{\alpha\beta}) = 
 \begin{pmatrix}
  |{\bm E}|^2 & -|{\bm E}|^2 & 0 & 0 \cr
  -|{\bm E}|^2 & |{\bm E}|^2 & 0 & 0 \cr
  0 & 0 & 0 & 0 \cr
  0 & 0 & 0 & 0 \cr
 \end{pmatrix},\nonumber
\end{equation}
where from null condition $|{\bm E}|=|{\bm B}| = E_y = B_z$. Comparison with (\ref{tensor_Ricci_Campo_nulo})
leads us to take $ |{\bm E}| =\epsilon$, such as:
\begin{equation}
 {\bm E} = (0,\epsilon,0) \hspace*{2cm} \mbox{and} \hspace*{2cm} {\bm B} = (0,0,\epsilon).\nonumber
\end{equation}
The potential 3-vector is ${\bm A}=(0,{\cal A},0)$, where ${\bm E}= -\dfrac{\partial {\bm A}}{\partial t}$ 
and ${\bm B}=\nabla \times {\bm A}$.

Take a 4-vector $A^{\alpha}$:
\begin{equation}
\label{potencial_A}
 A^{\alpha} = {\cal A}
 \begin{pmatrix}
  0 \cr 0 \cr 1 \cr 0
 \end{pmatrix} 
= {\cal A}\, v^{\alpha},
\end{equation}
where the vector $(v^{\alpha})= (0,0,1,0)$ has unit magnitude, $v_{\alpha} v^{\alpha} =1$ and it 
stands normal to $k^{\alpha}$,
\begin{equation}
 k_{\alpha} v^{\alpha} =0.
\end{equation}
Thus, we can write the electromagnetic field-strength,
\begin{equation}
\label{field_strenght_2}
 F_{\alpha\beta} =\epsilon\left( k_{\alpha} v_{\beta}-k_{\beta} v_{\alpha}\right),
\end{equation}
where it results in (\ref{field_strenght_1}),
\begin{equation}
\label{field_strenght_3}
  (F_{\alpha\beta})=
 \begin{pmatrix}
  0 & 0 & -\epsilon & 0 \cr
   0 & 0 & \epsilon & 0 \cr
   \epsilon & -\epsilon & 0 & 0 \cr
   0 & 0 & 0 & 0 \cr
\end{pmatrix}.
\end{equation}
We have that $F_{\alpha\beta}F^{\alpha\beta} =0$ and from (\ref{tensor_EM_F_2}) we have:
\begin{equation}
 T_{\alpha\beta} =  F_{\alpha\gamma}{F_{\beta}}^{\gamma} - 
 g_{\alpha\beta} \frac{1}{4} F_{\gamma\delta}F^{\gamma\delta} = \epsilon^2
 (k_{\alpha}v_{\gamma} - k_{\gamma}v_{\alpha})(k_{\beta}v^{\gamma} - k^{\gamma}v_{\beta}) =
 \epsilon^2\,k_{\alpha}k_{\beta}.
\end{equation}
A null electromagnetic field given by (\ref{field_strenght_2}) has only one null eigenvector $k^{\alpha}$ and no other.
The null electromagnetic field given by (\ref{field_strenght_2}) in the Minkowski frame 
is a null field in any other frame of reference.

We can return to Maxwell's equations in the Lorentz gauge (\ref{Maxwell_equations_4}) free of sources
\begin{equation}
\Box A_{\alpha}-  R_{\alpha\beta}A^{\beta} = 0, \nonumber
\end{equation}
where the term $  R_{\alpha\beta}A^{\beta} $, with aid of (\ref{potencial_A}) results in:
\begin{equation}
  R_{\alpha\beta}A^{\beta} = \kappa\epsilon^2\,{\cal A}\, k_{\alpha}k_{\beta}v^{\beta} = 0, \nonumber
\end{equation}
and we have:
\begin{equation}
\label{Maxwell_equations_6}
\Box A_{\alpha}  = 0. \nonumber
\end{equation}
In this situation we would expect to have solutions of Maxwell's equations of the form of oscillating waves.
The energy-momentum tensor (\ref{null_dust_tensor}),
 $T_{\alpha\beta} = \epsilon^2\, k_{\alpha}k_{\beta}$, may be considered as representing the incoherent superposition
 of waves with random phases and polarizations but the same propagation direction.

The term of ``pure radiation'' is used for an energy-momentum tensor,
\begin{equation}
\label{pura_radiacao}
 T_{\alpha\beta} = \epsilon^2\, k_{\alpha}k_{\beta}  \hspace*{1cm} \mbox{with}   \hspace*{1cm}
 k_{\alpha}k^{\alpha} = 0,
\end{equation} 
representing a situation in which all the energy is transported in one direction with the speed of light. 
Such tensors (\ref{pura_radiacao}) are also referred as null fields, null fluids and null dust.
This energy-momentum tensor (\ref{pura_radiacao}) also arises from other types of directed massless radiation,
for example, massless scalar fields, massless neutrino fields, or gravitational waves \cite{Stephani,Torre,Wils}.

\section{Conclusion}

The electromagnetic field-strength, $F_{\mu\nu}$, null or not, produces a Ricci curvature that satisfy the 
Rainich conditions (\ref{Rainich_conditions}). Conversely, any Ricci curvature tensor, null or not,
that satisfies the Rainich conditions (\ref{Rainich_conditions}), has a electromagnetic field tensor, $F_{\mu\nu}$,
that is unique up to a duality rotation.

For the situation where the electromagnetic field is general, non-null, and if the differential condition 
(\ref{curl_complexion}) is satisfied, one can find the complexion $\alpha$ up a additive constant by
integration (\ref{complexion_7}). With the complexion $\alpha$ one can choose an antisymmetric tensor $f_{\mu\nu}$,
with condition (\ref{condicao_extrema_2}),
 $f_{\mu\nu} \widetilde{f}^{\mu\nu}  = 0$, and by duality rotation (\ref{rotacao_dual_4}) one can 
 obtain $F_{\mu\nu}$, the antisymmetric tensor of electromagnetism, that obey Maxwell equations,
 $\nabla_{\mu}{\cal F}^{\mu\nu} =0$, without sources. We have seen if the electromagnetic field $F_{\mu\nu}$ has 
 a relationship with $f_{\mu\nu}$ by duality rotation (\ref{rotacao_dual_4}), the Ricci curvature tensor and 
 the energy-momentum tensor from electromagnetic field must be invariant under duality rotation of electromagnetic
 field, in accordance with (\ref{electrovac_3}). Then we have a way to express the vector 
 $\alpha_{\tau} = \partial_{\tau}\alpha$ (the gradient of the complexion $\alpha$ of the electromagnetic
 field) itself in terms of the Ricci curvature tensor in the form of equation (\ref{complexion_6}).
 The vector $\alpha_{\tau}$ contains the first derivative of the Ricci curvature tensor,
 and thus (\ref{curl_complexion})
 involves second derivative of the Ricci curvature. Consequently (\ref{curl_complexion}) contains up to the fourth
 order of the metric itself. This set of fourth order equations, plus the algebraic Raincih relations 
 (\ref{Rainich_conditions}), contains in the general case (${R_{\mu}}^{\rho}{R_{\rho}}^{\mu}> 0$) the entire
 content of Maxwell's equations for empty space plus Einstein's field equations with the Maxwell energy-momentum
 tensor as source. One set of equations of the fourth order (purely geometrical in character) takes the place of 
 two sets of equations of the second order. Misner and Wheeler called this resulting theoretical structure by
 ``already unified field theory''. This combined apparatus of gravitation theory and electromagnetism,
Misner and Wheeler named  ``geometrodynamics'' \cite{Misner}.

%





\bibliographystyle{elsarticle-num}
\bibliography{<your-bib-database>}







\end{document}